\documentclass[3p]{elsarticle}

\biboptions{sort&compress}
\journal{Acta Materialia}

\usepackage{amsmath}
\usepackage{mathtools}
\usepackage{setspace} 
\usepackage{float}
\usepackage{amssymb}

\usepackage{soul}

\usepackage{threeparttable}
\usepackage{etoolbox}
\usepackage{longtable}
\usepackage{multirow}
\usepackage{subfigure}    
\usepackage{booktabs}
\usepackage[utf8]{inputenc}
\usepackage{enumitem}
\usepackage{epsfig}

\usepackage[pdftex,dvipsnames]{xcolor}  

\usepackage{optidef}

\usepackage{hyperref}                 

\usepackage[colorinlistoftodos,prependcaption,textsize=tiny,textwidth=2cm]{todonotes}
\usepackage{xargs}                      

\newcommandx{\unsure}[2][1=]{\todo[linecolor=red,backgroundcolor=red!25,bordercolor=red,#1]{#2}}
\newcommandx{\info}[2][1=]{\todo[linecolor=OliveGreen,backgroundcolor=OliveGreen!25,bordercolor=OliveGreen,#1]{#2}}
\newcommandx{\improve}[2][1=]{\todo[linecolor=Plum,backgroundcolor=Plum!25,bordercolor=Plum,#1]{#2}}

\definecolor{darkgreen}{rgb}{0.1,0.8,0.1}

\newcommand{\insitu}{\textit{in-situ }}

\newcommand{\degree}{^{\circ}}

\newcommand{\us}{\textsubscript}

\newcommand\arsmall{$\sim 3 \times 3 \ $ µm}
\newcommand\arlarge{$\sim 13 \times 13 \ $ µm}

\newcommand{\cpm}{Enriched CP}
\newcommand{\cps}{Standard CP}

\newcommand{\SI}{SSLIP}

\soulregister\cite7
\soulregister\ref7
\soulregister\autoref7
\soulregister\us7
\soulregister\cpm7
\soulregister\cps7
\soulregister\degree7
\soulregister\textit7
\soulregister\textsubscript7
\soulregister\SI7

\begin{document}
\singlespace

\begin{frontmatter}

\title{Martensite plasticity and damage competition in dual-phase steel: A micromechanical experimental-numerical study}
%
%

\author[mymainaddress]{T. Vermeij \fnref{fn1}}
\address[mymainaddress]{Dept. of Mechanical Engineering, Eindhoven University of Technology, 5600MB Eindhoven, The Netherlands}
\author[mymainaddress]{C.J.A. Mornout \fnref{fn1}}
\author[mymainaddress]{V. Rezazadeh}
\author[mymainaddress]{J.P.M. Hoefnagels*}
\cortext[mycorrespondingauthor]{Corresponding author}
\fntext[fn1]{These authors should be regarded as joint first author.}
\ead{j.p.m.hoefnagels@tue.nl}

\begin{abstract}
Martensite damage in Dual-Phase (DP) steel has been studied extensively, yet, the exact deformation mechanisms that trigger or inhibit damage initiation remain mostly unexplored. Whereas generally assumed to be hard and brittle, lath martensite in fact deforms in a highly anisotropic manner, showing large strains under favorable habit plane orientations, which is attributed both to the lath morphology and to so-called 'substructure boundary sliding'. Yet, the correlation (or interplay) between plasticity and damage in lath martensite has not received much attention. Therefore, we raise the question whether these soft martensite plasticity mechanisms can delay or even inhibit damage initiation. We analyze several 'damage-sensitive' martensite notches, {i.e. thin contractions of two martensite islands,} by combining several state-of-the-art experimental and analysis methods. Deformations are tracked in-situ at the nanoscale, aligned to detailed microstructure maps, and categorised, for each martensite variant, into habit plane or out-of-habit-plane slip. In these experiments, strong plasticity ($>$70\%) is observed in martensite notches, enabled by slip along a favorably oriented habit plane, whereas damaged notches have unfavorably oriented habit planes, showing limited pre-damage strains ($<$10\%), carried by out-of-habit-plane slip. Additionally, one-to-one experimentally based Crystal Plasticity (CP) simulations are performed in parallel, employing a recently introduced Enriched CP approach which models a soft plasticity mechanism on the variants' habit plane. The Enriched CP simulations show considerably lower hydrostatic stresses in non-damaged and plastically deforming notches, thereby revealing that the soft habit plane mechanism is key for introducing the high plastic anisotropy that can lead to the inhibition of martensite damage in highly strained martensite notches. 
Finally, recommendations for improved damage inhibiting steels are proposed. 

DOI: \href{https://doi.org/10.1016/j.actamat.2023.119020}{doi: 10.1016/j.actamat.2023.119020}

\end{abstract}


\end{frontmatter}

\section*{Graphical Abstract}
\includegraphics[width=1\textwidth]{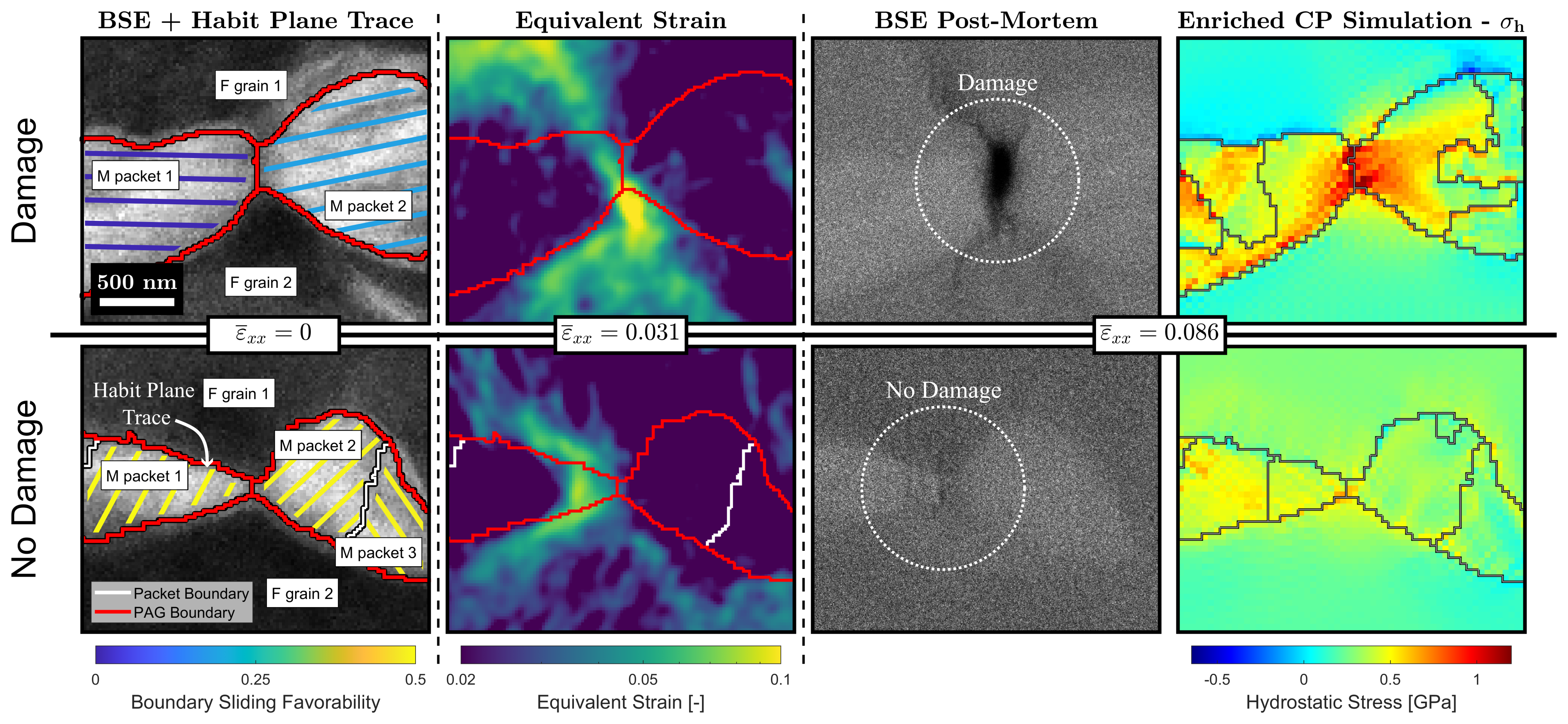}

\section{Introduction}

Advanced High Strength Steels (AHSS) combine favorable properties of different phases to optimize their mechanical properties \cite{Hilditch2015PropertiesAHSS}. Of the AHSS family, Dual-Phase (DP) steels, consisting of a soft ferrite (F) matrix with rather hard lath martensite (M) islands, are of interest because they combine a relatively low initial yielding, high strain hardening and high ultimate tensile strength whilst retaining relatively low production costs \cite{Tasan2015Review}. However, the large mechanical phase contrast between ferrite and martensite leads to strain partitioning and damage, of which the initiation and evolution have been analyzed extensively. Common damage mechanisms include "cracking" of martensite grains or notches {(thin contractions of two martensite islands)} \cite{AvramovicCingara2009a,hoefnagels2015retardation, archie2017micro}, void nucleation inside ferrite grains \cite{hoefnagels2015retardation, Ghadbeigi2013} and M-F interface damage \cite{AvramovicCingara2009a, hoefnagels2015retardation, archie2017micro,Ghadbeigi2013, liu2020revisiting}, of which the latter is dominant for lower martensite vol\% DP steel grades \cite{Lai2015a}. However, Hoefnagels \textit{et al.} highlighted the importance of the martensite island response in the onset of damage by reporting that damage initiates as brittle fracture of these islands, subsequently evolving in neighbouring ferrite, causing M-F interface damage \cite{hoefnagels2015retardation}. The likelihood for martensite damage is further reported to be enhanced by prior austenite grain boundaries (PAGBs) due to Mn embrittlement \cite{archie2017micro, Kuzmina2015GrainSteel}, as well as particular M-F notch geometries \cite{TASAN2014386, archie2017micro, Yan2015, deGeus2016a}.

Generally, martensite is considered to be hard and brittle. However, this view is strongly contrasted by recent investigations showing that, under certain conditions, martensite is able to deform in a ductile manner with high plastic strains, observed both in fully martensitic steels \cite{MORSDORF2016, Inoue2019, Du2016b} and in DP steels \cite{Yan2015, Ghadbeigi2013, Du2019, Ghadbeigi2010, VahidP4}, for which several explanations exist. Due to the lath morphology, dislocation glide parallel to the $(011)_{\alpha'}$ habit plane over two $\langle111\rangle_{\alpha'}$ slip directions is less obstructed as compared to slip over other planes, on which dislocation movement is restricted by internal substructure boundaries, resulting in anisotropic behavior \cite{MICHIUCHI2009, mine2013micro,Du2016a, Ungar2017CompositeSteels, Harjo2017WorkAnalysis}. Additionally, strong plasticity can be enabled by sliding of laths over their boundaries, termed as substructure boundary sliding (SBS) \cite{Inoue2019, MORSDORF2016, Du2016b, Du2019}. SBS can occur over all boundaries inside a packet and activates at a lower critical resolved shear stress (CRSS) as compared to regular crystallographic slip \cite{Du2019}, which further aids in explaining the large anisotropy observed in martensite plasticity \cite{MORSDORF2016, MICHIUCHI2009, kwak2016anisotropy}. The deformation mechanism behind SBS has been attributed to 'greasy' sliding along thin films of retained austenite between the laths \cite{Maresca2014a, maresca2016reduced}, which can easily deform by slip along the $\{111\}_{\gamma}$ plane, oriented parallel to the laths' $(011)_{\alpha'}$ habit plane \cite{Morito2003}. Alternatively, recent numerical investigations suggest that the transformation of austenite to martensite upon applied stress can enable enhanced plasticity along the habit plane \cite{Maresca2018}, which agrees with observations that retained austenite films disappear upon deformation \cite{Inoue2019, MORSDORF2016}. Additionally, boundary sliding may be explained by the common $\{011\}\langle111\rangle_{\alpha'}$ slip systems of martensite variants of the same block, which would enable dislocation glide along the block boundary interface \cite{MORSDORF2016, Ohmura2004Dislocation-grainMicroscope}. Sliding of lath boundaries has been observed experimentally in micro-tensile samples \cite{Du2016b, Du2019}, in fully martensitic steel \cite{Inoue2019} and likely occurs in DP steels as well \cite{MORSDORF2016,Du2019, liu2020revisiting,Tian2020, vermeij2022nanomechanical, VahidP4}. In DP steel, single-packet PAGs are common, which increases anisotropy and may locally enhance SBS, since lath boundaries are more likely to cross the complete width of a martensite island \cite{Du2018a}. 

These observations of large martensite plasticity appear to be in contrast with many reports of early martensite damage initiation (see e.g. Refs.  \cite{Materkowski1979TemperedSteel, Bowen1984EffectsSteel, Krauss1999, Horn1978MechanismsSteels}) claiming to be of a brittle nature, posing the question whether there is a correlation between large plastic deformation and damage initiation, or the inhibition of damage, in martensite. Tasan \textit{et al.} analyzed the global development of plasticity until damage initiation and concluded that M-F interface damage occurs at boundaries of highly deformed zones \cite{TASAN2014386}. Ghadbeigi \textit{et al.} proposed that cracks initiate from the M-F interface regions, followed by propagation towards the interior of martensite islands \cite{Ghadbeigi2013}. Yan \textit{et al.} analyzed the ultimate strain levels before damage for various M-F geometric morphologies \cite{Yan2015}, yet provide no insights into the mechanisms. Moreover, in most works, e.g. by Hoefnagels \textit{et al.} \cite{hoefnagels2015retardation}, the focus is solely on areas where damage occurred, whereby strongly plastically deforming areas are mostly not analyzed.

Full analyses of specific deformation mechanisms that occur before martensite damage initiation are very rare in the literature. More importantly, the apparent contradiction between observations of early damage initiation and extreme plasticity in martensite, possibly induced by SBS, has not been investigated. Therefore, we aim to analyze the competition between damage initiation and large plasticity in martensite notches, which are known to be sensitive to damage \cite{deGeus2016a}, in DP steel by monitoring the nanoscale deformation mechanisms that may or may not lead to damage, with a prime focus on the orientation of the habit plane. {Additionally, recent developments in integrated experimental-numerical testing have yielded deeper insights in the mechanical behaviour of a range of steels and alloys \cite{TASAN2014386, cho2013three, Tasan2014, benzing2019experimental, yanagimoto2019contribution, motaman2020anisotropic, molter2021role, el2022multi}. This enables next generation materials design by unraveling of the often complex micro-deformation mechanisms that govern the macroscopic material behavior.}

To analyse the plasticity and damage competition in martensite islands, a range of experimental and numerical techniques with high levels of detail are employed. Micrometer sized regions of interest (ROIs) are selected by identification of damage-sensitive martensite notches, based on the prior identification of damage hotspots, achieved by extending the damage hotspot statistics methodology of de Geus \textit{et al.} \cite{deGeus2016a}. After careful microstructural characterization, ultra-high-resolution Scanning Electron Microscopy based Digital Image Correlation (SEM-DIC) is employed to obtain nanoscale deformation fields, which are aligned to the microstructure using a recently developed alignment procedure \cite{vermeij2022nanomechanical, Vermeij2021}. Novel grain and PAG reconstruction approaches \cite{hielscher2022variant} allow for deformation analysis at the level of individual grains, which includes both conventional Schmid factor (SF) analyses and employment of a recently proposed Slip System based Local Identification of Plasticity (SSLIP) method to determine active slip system activity maps, based on deformation fields from DIC \cite{vermeij2022crystallographic}. Another important link in understanding damage initiation is the stress state in the material, which can be estimated by means of crystal plasticity (CP) simulations of the experimental microstructure \cite{TASAN2014386, Diehl2017CrystalMicrostructure}. Recent works have successfully captured the experimentally observed Substructure Boundary Sliding with new CP modelling methods \cite{kwak2016anisotropy, Maresca2014a, maresca2016reduced, liu2020revisiting, Rezazadeh2022AnSubstructure, VahidP4}, of which one will be employed in this work \cite{VahidP4}. All these advanced experimental-numerical methodologies are combined and integrated to analyze in detail several ROIs in a DP600 specimen, deformed globally under plane strain tension, containing damage-sensitive martensite notches that show either damage or large plastic deformation.

\section{Methodology}
\label{sec:method}

\subsection{Experimental Methodology}

\begin{figure}[H]
    \centering
    \includegraphics[width=0.85\linewidth]{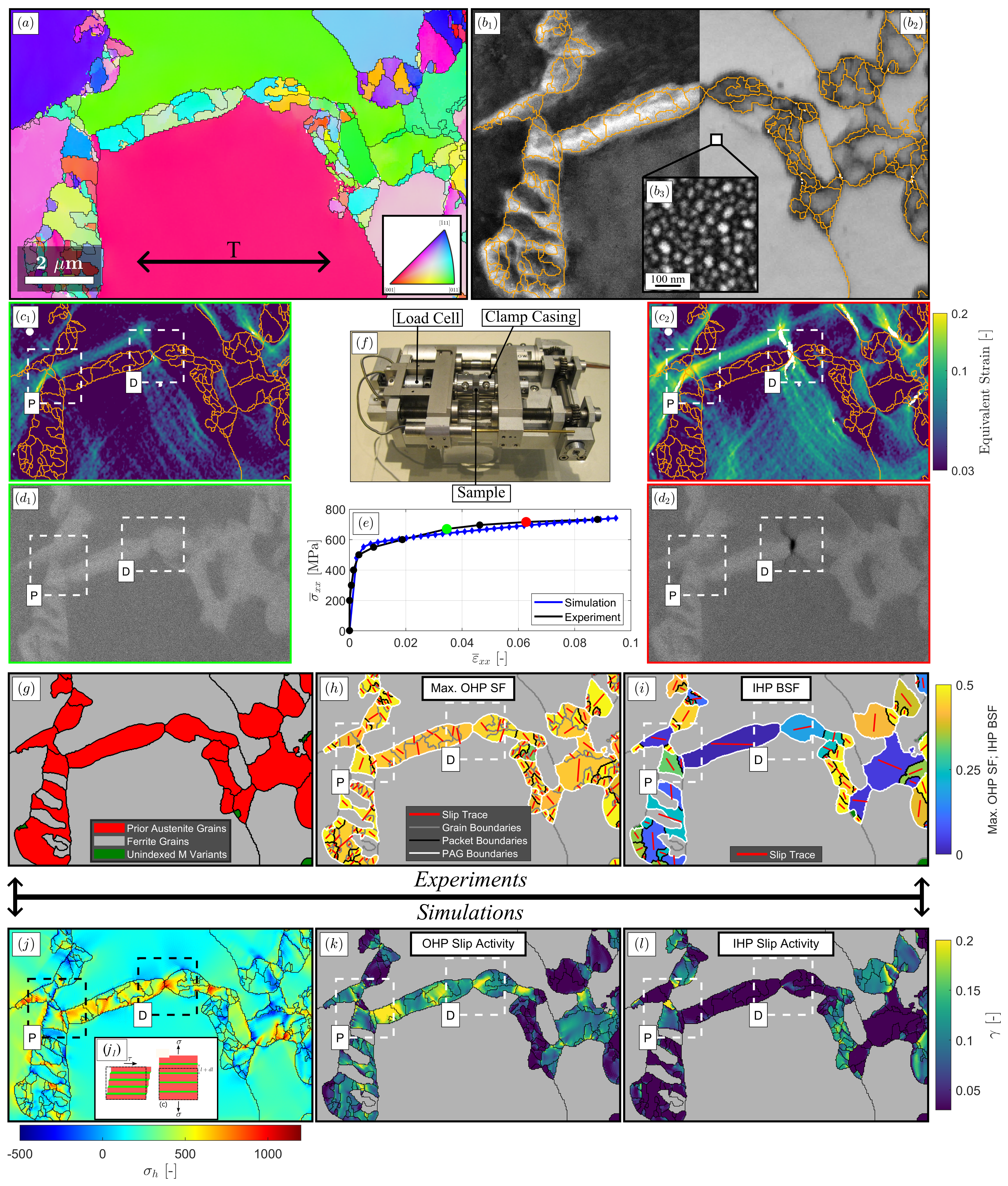}

    \caption{{Overview of experimental and numerical data obtained for one region.} (a) IPF map including grain boundaries, which are also plotted in other aligned maps. Direction of globally applied plane strain tension is indicated by a "T". (b\textsubscript{1}) ECCI and (b\textsubscript{2}) EBSD CI maps. An inset of the DIC pattern is shown in (b\textsubscript{3}). (c\textsubscript{1}) Equivalent strain map (in reference configuration) and (d\textsubscript{1}) in-lens BSE image (in deformed configuration) of an early deformation increment. Indicated are two ROIs, one (ultimately) showing damage (D) (see \autoref{fig:Paper_DROI1} for details) and the other showing large plasticity without damage (P) (see \autoref{fig:Paper_PROI2} for details). (c\textsubscript{2}) Equivalent strain and (d\textsubscript{2}) in-lens BSE maps for a later deformation increment. (e) Global stress-strain curve of the experiment (black) {and from a matching simulation using the \cpm\  model (blue). The green and red increments correspond to subplots (c\textsubscript{1}/d\textsubscript{1}) and (c\textsubscript{2}/d\textsubscript{2}), respectively. (f) Photo of the micro-tensile stage with the clamped sample. 
    (g,h,i) Results of Prior Austenite Grain (PAG) reconstruction. (g) Overview of prior austenite grains and unindexed M variants. (h) The Max. out-of-habit-plane (OHP) Schmid Factor (SF) map, including the slip plane trace of the OHP slip system with highest SF for each variant. (i) In-habit-plane (IHP) boundary sliding favorability (BSF) for each packet, including the habit plane trace. (j,k,l) Overview of the simulation results using the \cpm\ model. (j) Simulated hydrostatic stress field.     (j\us{2}) Schematic view of the "matrix-film" (martensite-austenite) model which is represented in each martensite voxel. (k) Sum of the simulated OHP martensite slip activity. (l) sum of the simulated IHP martensite slip activity.}
    }    
    \label{fig:fig1method}
\end{figure}

DP600 with a $\sim20 \%$ lath martensite volume fraction with composition 0.092C-1.68Mn-0.24Si-0.57Cr wt\% was given a minor heat treatment of 10 minutes at 750$\degree$, followed by water quenching, in order to retrieve a true dual-phase, ferritic-martensitic, steel microstructure without pearlite and bainite \cite{Du2018a}.  A specimen (sized $25\times5\times0.5$ mm) was metallographically prepared up to a final mechanical polish with OPS colloidal silica particles. Multiple \arlarge\ regions with damage-sensitive martensite notches were selected by predicting the likelihood of (future) damage development by means of a novel damage prediction method \cite{Wijnen2022Damage}, which is based on the damage hotspot statistics methodology of de Geus \textit{et al.} \cite{deGeus2016a}. For each region, see for example \autoref{fig:fig1method}, Electron Backscatter Diffraction (EBSD) scans (Edax Digiview 2 camera on a Tescan Mira 3 SEM), with step size 30 nm, were obtained before deformation and processed using EMSphinx \cite{LENTHE2019112841}, which improves the indexing of martensite variants considerably. Electron Channeling Contrast Imaging (ECCI) scans were made for additional information on the phase distribution. \autoref{fig:fig1method} shows, among others, (a) an Inverse Pole Figure (IPF) map of an EBSD scan, (b\textsubscript{1}) an ECCI scan and (b\textsubscript{2}) a Confidence Index (CI) map (from EMSphinx). The specimen was deformed in situ by using a micro-tensile stage (Kammrath \& Weiss) inside a Tescan Mira 3 SEM, approaching plane strain tension by using a 6:5 width-length ratio for the non-clamped {(millimeter sized)} sample region, to prevent necking, see \autoref{fig:fig1method}(f). Deformations are tracked using high-resolution SEM-DIC. An InSn DIC speckle pattern is applied by employing the one-step patterning method introduced by Hoefnagels \textit{et al.} \cite{HoefnagelsPattern}, using the following InSn sputter coating parameters: $9\times10^{-3}$ mTorr chamber pressure, 20 mA current, 2 minutes sputtering time, 90 mm target-to-sample distance, resulting in a $\sim$ 20-50 nm speckle size, see \autoref{fig:fig1method}(b\us{3}). 
All (inlens) SE/BSE images acquired were obtained with the following settings: 7 mm working distance, pixel size 5 nm, region sizes ranging between $2048\times2048$ and $3072\times3072$ pixels, electron voltage either 5kV ("low-eV") or 20 kV ("high-eV"). 
After patterning, but before applying deformation, a high-eV SE/BSE image, which shows contrast of both the InSn DIC pattern and the underlying microstructure, of each of the regions is acquired, which is used for precise alignment of the EBSD and ECCI microstructure datasets to the deformation fields. During the \insitu test, in-lens, low-eV SE images, which only show the InSn DIC pattern contrast, are obtained in ten successive increments up to $\sim 9 \%$ global $\overline{\varepsilon}_{xx}$. High-eV in-lens BSE images are taken at various increments to objectively identify the initiation of damage. DIC is performed on the low-eV SE images using MatchID (subset size 21 pixels, step size 1 pixel), employing DIC and strain calculations strategies as described in previous work \cite{vermeij2022nanomechanical, Vermeij2021}. The equivalent 2D von Mises strain \cite{vermeij2022nanomechanical}, based on the true strain tensor, is plotted in all figures in this work, unless mentioned otherwise. 

Data alignment is performed using tools from the nanomechanical testing framework as proposed by Vermeij \textit{et al.} \cite{vermeij2022nanomechanical} \footnote{{While we use the DIC parameters and data alignment tools from "the nanomechanical testing framework" referenced here, the experiments conducted in the current manuscript do not consist of nano-tensile tests, but consist of millimeter scale samples on which micrometer sized areas are tracked.}}, resulting in microstructure-resolved strain fields on a grid with a pixel size of 20 nm, which is just below the spatial resolution of DIC and EBSD. The alignment can be observed, e.g., in the two strain increments, plotted in the reference configuration, in \autoref{fig:fig1method}(c\textsubscript{1}/c\textsubscript{2}), of which the increments are indicated in the global stress-stain curve in (e). The grain boundaries, calculated from the EBSD data and plotted over the other fields, showcases the quality of the alignment between the datasets. In-lens BSE images of the two deformation increments, in the deformed configuration and thus not aligned, are displayed in (d\textsubscript{1}/d\textsubscript{2}) to monitor damage initiation and evolution. The direction of tension is indicated by a "T" in \autoref{fig:fig1method}(a).

\subsection{Analysis Strategies for Identification of Martensite Deformation}
\subsubsection{EBSD Analysis and PAG Reconstruction}

All EBSD and deformation data is further processed using the MTEX toolbox in MATLAB \cite{bachmann2010, MTEX}. Martensite and ferrite are identified by thresholding the CI data. Inspection of the grain boundary overlay on the ECCI/CI scan in \autoref{fig:fig1method}(b\textsubscript{1,2}) reveals a clear mismatch between the positions of the M-F interfaces, with the ECCI/CI data systematically showing larger martensite islands. {This is problematic as it may hinder the proper allocation of plasticity to ferrite or martensite. The mismatch is caused by (i) the limited spatial resolution of EBSD (estimated to be $\sim 100$ nm) and (ii) the higher quality of ferrite EBSD patterns in comparison to martensite patterns, which causes martensite to be indexed as ferrite at or near the M-F interface. To alleviate this problem, we apply a dedicated grain reconstruction and extension routine, as described in \ref{sec:App_recon}, with the resulting extended grains shown in \autoref{fig:fig1method}(g-l).
}

{PAG reconstruction is required for determination of the habit plane orientation of each variant, which is crucial for the analysis of plasticity and damage in this paper. Here, we apply the Variant Graph method for PAG reconstruction as implemented in MTEX \cite{Nyyssonen2016IterativeMisorientations, hielscher2022variant, niessen2022parent}, with several extra steps in order to handle the small martensite islands which have a limited number of variants. See \ref{sec:App_pag} for a detailed explanation. The PAG reconstruction results in a prior austenite grain map as shown in \autoref{fig:fig1method}(h). Each PAG consists of a maximum of four packets, each of which contains martensite variants which share a common habit plane.}

\subsubsection{Schmid Factor analysis}
\label{sec:IdPlasticity}

With the variants and habit plane orientations identified, a Schmid Factor (SF) analysis is first performed to calculate the resolved shear stress for each slip system, with respect to the global uniaxial tension stress. The standard equation $SF=cos(\phi)\cdot cos(\lambda)$ is employed, where $\phi$ is the angle between slip plane normal and the direction of tension and $\lambda$ is the angle between slip direction and the direction of tension. Although martensite formally has a Body-Centered Tetragonal (BCT) lattice structure, its tetragonality is very low due to the low C content \cite{lobodyuk2019tetragonality}, which is why a Body-Centered Cubic (BCC) lattice structure is assumed, which is common in the literature \cite{Mohammed2018MultiscaleSteel, Tjahjanto2015MultiscaleScheme}. In this work, we distinguish between out-of-habit-plane (OHP) and in-habit-plane (IHP) slip. For OHP slip, the maximum SF is found among the 10 $\{110\}\langle111\rangle_{\alpha'}$ BCC slip systems whose slip plane is not parallel to the habit plane, which is called 'Max. OHP SF'. For IHP slip, the concept of "Boundary Sliding Favorability" (BSF) is introduced, which is based on the maximum resolved shear stress direction along the $(011)_{\alpha'}$ habit plane and is calculated as $BSF=cos(\phi)\cdot sin(\phi)$ \cite{VahidP4}. The BSF approximates activation by means of Substructure Boundary Sliding (SBS), which may occur over various in-habit-plane directions, caused by different mechanisms as explained in the introduction (i.e. $\{111\}\langle110\rangle_{\gamma}$ FCC slip systems, regular BCC slip inside the lath, boundary sliding on the $\{110\}_{\alpha'}$ interfacial planes, or other habit-plane aligned plasticity mechanisms). In \autoref{fig:fig1method}(g,i), the magnitude of the Max. OHP SF and the BSF for IHP slip are shown, respectively, including corresponding slip plane traces. Packets share a habit plane and thus have the same BSF. Note that the few unindexed martensite grains, which are not assigned to a PAG, have no known habit plane BSF.

\subsubsection{Local slip system identification}
{While the SF and BSF indicate if a slip system is likely to activate, analysis of the DIC data actually allows the direct identification of active slip systems. Here, we employ a very recent, novel methodology: Slip System based Local Identification of Plasticity (\SI) \cite{vermeij2022crystallographic}. \SI\ identification uses the measured 2D displacement gradient tensor to estimate the most likely (combination of) active slip systems for each individual data point in the deformation map, by solving an optimization problem to obtain all slip system amplitudes. As explained and demonstrated in more detail in \cite{vermeij2022crystallographic}, it yields a slip activity map for every considered slip system in a certain grain. In this work, for both types of plasticity, i.e. OHP and IHP, we sum all the respective slip system activities at each position, which results in an OHP and IHP slip activity map, giving direct insights into the role of the habit plane on the plasticity. \ref{sec:App_sslip} explains this in more details and showcases the method on a small area from \autoref{fig:fig1method}.}

\subsection{Standard \& Enriched Crystal Plasticity simulations}

Along with the extensive analysis of experimental data, the use of numerical simulations is vital for gaining new insights into the plasticity and damage mechanisms of lath martensite. It allows for generation of fields that are difficult to obtain in experiments, such as stress fields and contributions of individual slip systems, of which the latter can be compared to results found with the \SI\ method.

In this work, Crystal Plasticity (CP) modelling is used to simulate the deformations in ferrite and martensite, with the aim of matching the experiments, thereby allowing the analysis of local stresses using the simulated results. Through the use of the Düsseldorf Advanced Material Simulation Kit (DAMASK), phenomenological CP is employed in martensite and ferrite, employing 12 $\{110\}\langle111\rangle$ slip {systems for both phases. On top of this \cps, we also explore a more advanced model termed as the Enriched CP approach, which was recently developed by Rezazadeh \textit{et al.} \cite{Rezazadeh2022AnSubstructure, VahidP4}, based on the reduced model by Maresca \textit{et al.} \cite{maresca2016reduced}. The proposed model is a Standard CP model described in a large deformation formulation \cite{Roters2019}, but with the difference that the habit plane of each martensite variant is enriched by 3 soft FCC slip systems, which have a Critical Resolved Shear Stress (CRSS) of half that of the BCC martensite. These FCC slip systems are positioned according to the Kurdjumov-Sachs (KS) Orientation Relationship (OR), to represent slip in the thin films of retained austenite, i.e. substructure boundary sliding. The plasticity in the other 9 slip systems of the FCC crystal is excluded in line with the characteristic morphology of lath martensite. Therefore, the plastic contribution of the retained austenite film is limited to the IHP slip systems only. It is important to mention that, by modelling thin austenite films with 3 easily activated IHP slip systems whose slip directions are rotated $60 \degree$ with regard to each other, the \cpm\ model can (implicitly) account for any of the soft mechanisms (as mentioned in the introduction) parallel to the habit plane of the martensite laths that may be active in reality. More details on the simulation framework can be found in \ref{sec:app_sim} and the interested reader is referred to the work of Rezazadeh \textit{et al.} \cite{VahidP4}.

To construct the model, the obtained experimental microstructure is taken as input into the simulation framework, allowed by the use of the voxel-based grid structure of DAMASK. Aligned EBSD data points, containing crystal orientations and phases, are used directly in the simulation and can be complemented by the habit plane orientations of each variant \cite{VahidP4} when employing the \cpm\ model. In that case, each martensite voxel contains the reduced martensite-austenite laminate model, as visualized schematically in \autoref{fig:fig1method}(k\us{2}) \cite{VahidP4}.

Upon loading of the \cpm\ simulation using boundary conditions that closely resemble the experiments, deformation data can be extracted in the form of summed simulated slip activities and stress fields, similiar to how the \SI\ results are summed. \autoref{fig:fig1method}(j) shows the total simulated OHP slip activity map, \autoref{fig:fig1method}(k) shows the simulated hydrostatic stress map and \autoref{fig:fig1method}(l) shows the total simulated IHP activity map. The slip activity fields will be compared to the experiments, for verification, while the hydrostatic stress fields will serve to indicate the likelihood for damage. Finally, the global stress-strain behaviour of this particular area, using the \cpm\ model, is added to \autoref{fig:fig1method}(e), as a blue line, where it agrees well with the experimental curve.

The Enriched CP model, the simulation strategy, the material parameters and the boundary conditions are described in more detail in \ref{sec:app_sim}.
}

\section{Results and Discussion}

Five small \arsmall\ regions of interest (ROIs) are investigated, containing damage-sensitive martensite notches that either show damage with some preceding plasticity or large plastic deformation without damage. Martensite notches are defined as contractions of two martensite islands of less than 500 nm wide. They are susceptible to damage due to their geometry and contrast with the ferrite phase \cite{TASAN2014386, Yan2015, deGeus2016a}. Several other notches were analyzed in detail, but were discarded from further analysis because of any of the following reasons: inconclusive phase distribution, non-unique or poorly fitting PAG orientations, poorly indexed EBSD data, inconclusive slip system identification, and/or no occurrence of either local plastic deformation or damage. Nonetheless, the five remaining ROIs, each with high quality data of an interesting martensite notch, could be analyzed in great detail, as discussed next. 

\begin{figure}[H]
    \centering
    \includegraphics[width=\linewidth]{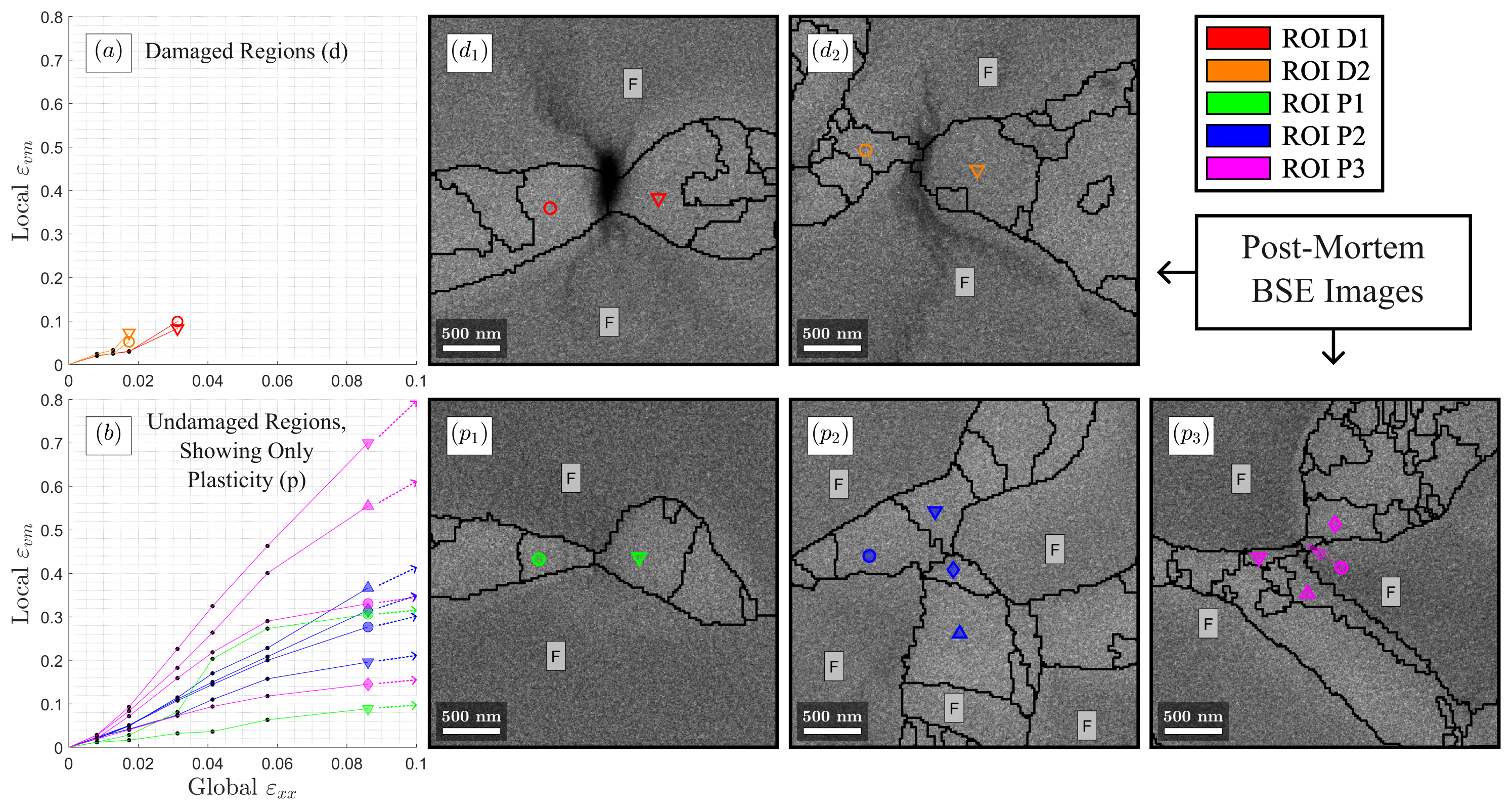}
    \caption{Local deformation behavior of martensite notches. (a,b) relation between global $\overline{\varepsilon}_{xx}$ and local maximum measured strain $\varepsilon_{vm}$ for individual variants, in (a) damaged and (b) non-damaged regions. The arrows in (b) indicate that plasticity may increase further upon continued deformation. In-lens BSE images (post-mortem) are displayed for two ROIs that show damage (d\textsubscript{1-2}) and for three ROIs that deformed strongly without damage (p\textsubscript{1-3}). The grain boundaries, identified from the EBSD data and forward deformed by the displacement fields from DIC \cite{vermeij2022nanomechanical}, are plotted on top of each BSE image. Ferrite grains are denoted by F. The colored symbols in the martensite variants correspond to the colored symbols used in (a) and (b). The ROIs shown in (d\textsubscript{1}) and (p\textsubscript{2}) are taken from the larger region displayed in \autoref{fig:fig1method} \& \autoref{fig:PAGrecon}.
    }
    \label{fig:eqStrains}
\end{figure}

The five ROIs are shown in \autoref{fig:eqStrains}, where (a) and (b) display, for respectively damaged and undamaged regions, the highest reported equivalent strain in each deformation increment, for relevant variants, plotted against the global average $\overline{\varepsilon}_{xx}$. For each ROI, a post-mortem in-lens BSE image is shown in which voids can clearly be identified as dark spots, while strong plastic localization bands (mostly in Ferrite) show up as a slightly darker grey. The ROIs in (d\textsubscript{1-2}) show damage, whereas those in (p\textsubscript{1-3}) deform strongly without damage. Grain boundaries, obtained from EBSD data combined with subsequent grain extension, are forward deformed using DIC displacement fields and are aligned to the post-mortem BSE images \cite{vermeij2022nanomechanical}. For the damaged regions, strains are reported until damage initiated, which was observed through the in-situ BSE images during deformation. In the final increments of deformation, DIC performs less well due to large local plastic deformations, which results in uncorrelated regions inside the displacement fields. Therefore, in the case of large plasticity in the absence of damage, these displacement fields are interpolated with cubic shape functions, resulting in an underestimation of the peak strain in the plasticity band. The difference in strain carrying capacity between the two types of notches is large, with the two damaged notches deforming to less than $10\%$ strain, whereas the three non-damaged notches deform at least to $\sim 31 \%$, $\sim 37 \%$ and $\sim 70 \%$ strain. Here, it is important to realize that (i) the strain in these notches is underestimated and (ii) these notches could also have deformed further, as indicated by the arrows in \autoref{fig:eqStrains}(b).

A detailed overview of the five \arsmall\ ROI is displayed in \autoref{fig:Paper_DROI1}, \ref{fig:Paper_DROI2}, \ref{fig:Paper_PROI3}, \ref{fig:Paper_PROI2} and \ref{fig:Paper_PROI3}, each presented in the same layout. Subplot (a) shows the (extended) grain boundaries, overlaid on an ECCI image to highlight the phase distribution. Subplots (b\textsubscript{1,2}) display the maximum out-of-habit-plane (OHP) SF and the in-habit-plane (IHP) Boundary Sliding Favorability (BSF) maps. Subplots (c\textsubscript{1,2}) visualize the occurrence or absence of damage: (c\textsubscript{1}) maps the equivalent strain at a late increment of deformation where, for non-damaged ROIs, the occasional small gaps in the displacement fields are interpolated, while for damaged ROIs, the uncorrelated gaps in the strain maps are too large to enable proper interpolation. Subplot (c\textsubscript{2}) shows the in-lens BSE image at the final experimental strain increment, displayed without forward-deformed grain boundaries (as was already shown in \autoref{fig:eqStrains}) to give a clearer view of the damage location. The equivalent strain map of the first increment of noticeable strain plasticity is shown in (d). The OHP- and IHP-summed slip activity (obtained through the \SI\ method, see \autoref{fig:sslip}) corresponding to this deformation increment is shown in (e\textsubscript{1,2}). The numerical results for both \cpm\ and \cps\ models are shown in subplots (f) and (g). Note that the grain boundaries are coarser in these plots because of the coarsening of the spatial resolution from 20 to 40 nm. For each model, the following data is presented: (1) $\varepsilon_{vm}$ at the same $\overline{\varepsilon}_{xx}$ as the experimental increment in (d), (2) the sum of OHP slip activity at this increment, (3) the sum of IHP slip activity at this increment, and (4) the hydrostatic stress at the same $\overline{\varepsilon}_{xx}$ as the final experimental increment. All simulations were performed on regions which are approximately 20 times larger (\arlarge) than the ROIs. In Figures 6-10, only the simulation results for the smaller ROIs are shown.

\subsection{ROIs with Martensite damage}
\label{ROIdamage}
\begin{figure}[H]
    \centering
    \includegraphics[width=1\linewidth]{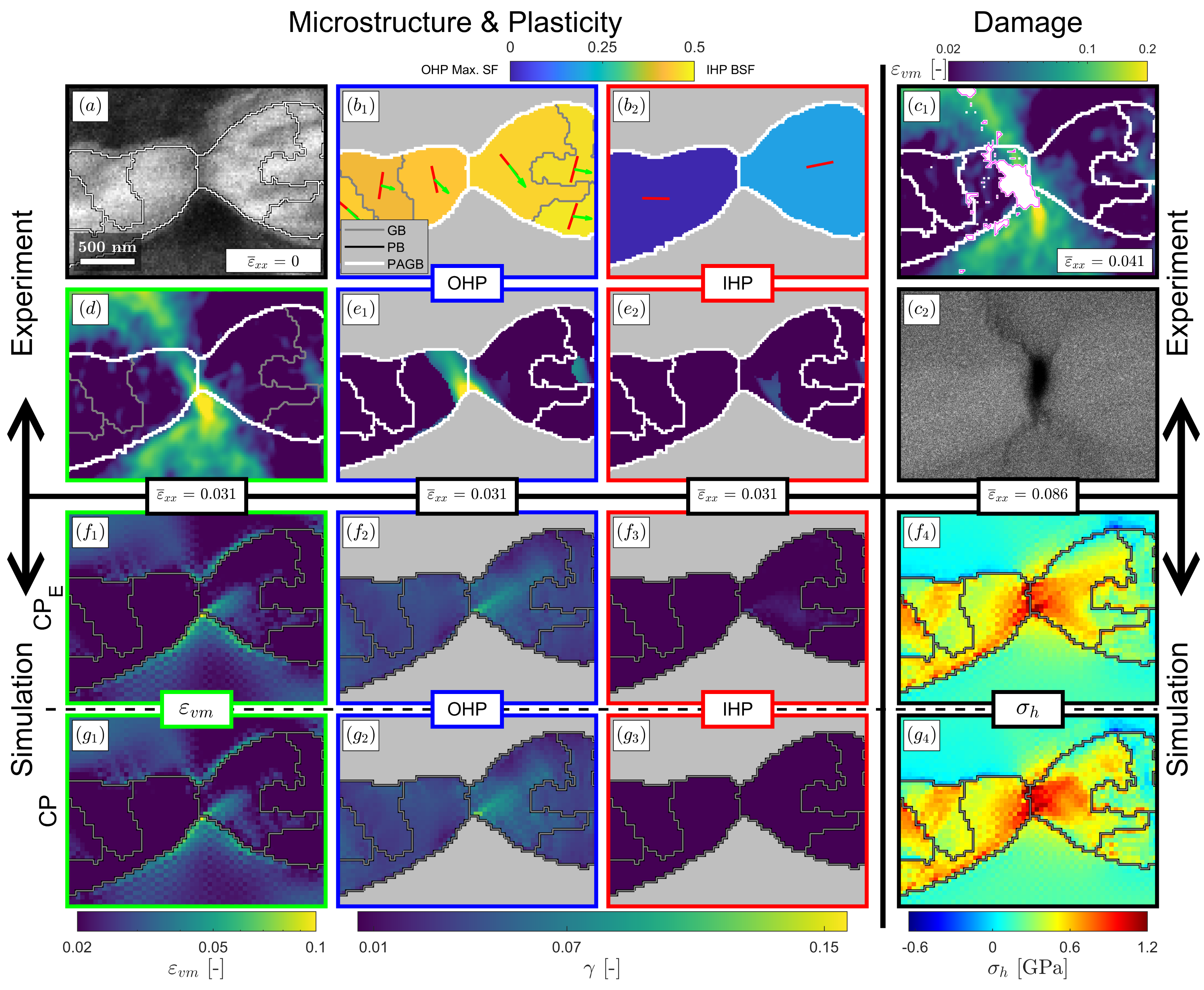}
    \caption{Overview of experimental-numerical results for the \arsmall\ ROI D1. (a) Undeformed microstructure with extended grain boundaries, plotted on top of an ECCI image. (b\textsubscript{1},b\textsubscript{2}) Maximum out-of-habit-plane Schmid factor (Max. OHP SF) and in-habit-plane boundary sliding favorability (IHP BSF) maps including their slip traces in red (and slip directions in green for OHP slip). Grain, packet and PAGBs are shown. (c\textsubscript{1}) Equivalent strain map for the first increment after damage initiation, in which uncorrelated regions are indicated by white areas with pink boundaries. (c\textsubscript{2}) In-lens BSE image at the final increment of deformation (not aligned, the aligned image is shown in \autoref{fig:eqStrains}(d\textsubscript{1})). The global $\overline{\varepsilon}_{xx}$ is indicated for both (c\textsubscript{1}) and (c\textsubscript{2}). (d) Equivalent strain map for the first increment of noticeable localization, for which the global $\overline{\varepsilon}_{xx}$ is indicated as well. (e\textsubscript{1},e\textsubscript{2}) Summed experimental OHP and IHP slip activity, based on \SI, corresponding to the strain map in (d). (f,g) Simulation results: (f\textsubscript{1},g\textsubscript{1}) Equivalent strain, (f\textsubscript{2},g\textsubscript{2}) summed OHP slip activity, (f\textsubscript{3},g\textsubscript{3}) summed IHP slip activity and (f\textsubscript{4},g\textsubscript{4}) Hydrostatic stress. The global $\overline{\varepsilon}_{xx}$ for (e\textsubscript{1,2}), (f\textsubscript{1,2,3}) and (g\textsubscript{1,2,3}) is the same as for the equivalent strain map shown in (d), the global $\overline{\varepsilon}_{xx}$ for (f\textsubscript{4}) and (g\textsubscript{4}) is the same as for (c\textsubscript{2}). The color scales for 2D equivalent strain (logarithmic, column 1) and slip activity (linear, column 2-3) are the same for experimental and numerical fields. The equivalent strain field in (c\textsubscript{1}) has an individual color scale.}
    \label{fig:Paper_DROI1}
\end{figure}

\begin{figure}[H]
    \centering
    \includegraphics[width=0.95\linewidth]{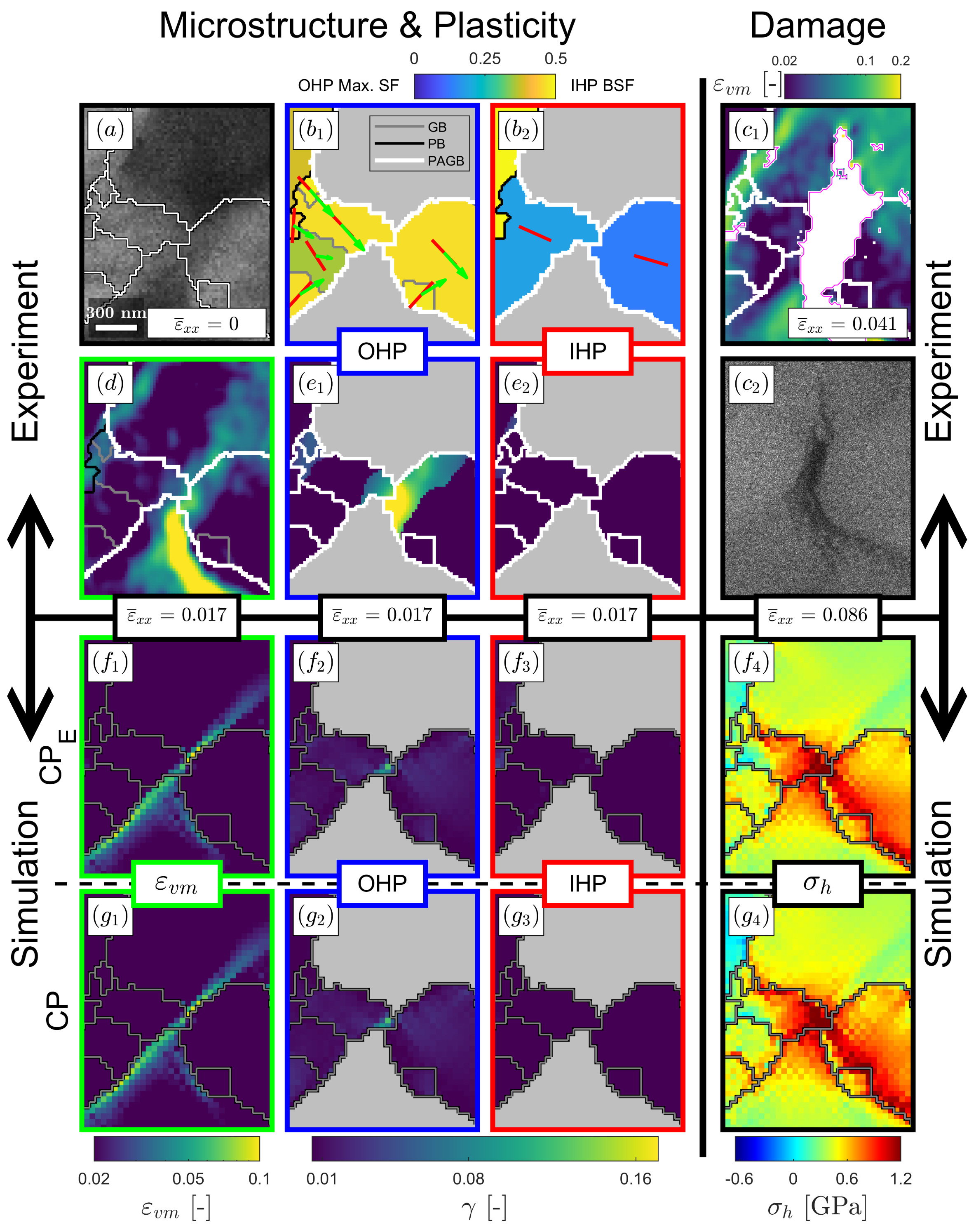}
    \caption{Overview of experimental-numerical results for ROI D2. Subplots are described in the caption of \autoref{fig:Paper_DROI1}.}
    \label{fig:Paper_DROI2}
\end{figure}

First, the two damaged martensite notches are discussed: ROI D1 (\autoref{fig:Paper_DROI1}) and ROI D2 (\autoref{fig:Paper_DROI2}). The damage initiates as a micro-crack at the thinnest part of the notch, which then opens to form a void, after which damage further extends into the ferrite phase, an evolution that is similar to what was reported by Hoefnagels \textit{et al.} \cite{hoefnagels2015retardation}. Both notches form a horizontal connection of two packets of different PAGs, which all have unfavorably oriented habit planes (i.e. BSF $<0.2$), whereas the maximum SF for OHP slip is much higher, close to 0.5 (compare subplots (b\textsubscript{1}) vs. (b\textsubscript{2})). The horizontal geometry of the notches with regard to the direction of tension make them damage-sensitive \cite{archie2017micro, Yan2015, deGeus2016a}. Additionally, damage is hypothesized to be more likely along PAG boundaries (PAGBs) because of additional Mn embrittlement \cite{archie2017micro}. The observed maximum strains of $7-10\%$ before damage (\autoref{fig:eqStrains}) coincide with values from the work of Yan \textit{et al.} \cite{Yan2015}, in which maximum strains of $\sim 5-15 \%$ for similar notch geometries with a notch thickness of $\sim 300 \ nm$ were reported. Note that, however, both for ROI D1 and D2, the strain localization does not occur exactly over the PAGB at the thinnest part of the geometry, see subplot (e\textsubscript{1}). 
\SI\ investigation shows that almost all plastic deformation is identified as OHP slip. Almost no habit plane activity is noted, likely due to a particularly unfavorable habit plane orientation in the packets that neighbour the notch, see subplot (e\textsubscript{2}). Despite the low IHP BSF, there is still some IHP activity on the right packet of the notch in ROI D1, but this is insufficient to prevent damage in the adjacent packet.

Simulations of ROI D1 show that strains in the notch do not reach the same maximum level. In the experiment, the maximum strain is $10\%$ (\autoref{fig:Paper_DROI1}(d)) compared to maximum $5\%$ in the simulations (\autoref{fig:Paper_DROI1}(f\textsubscript{1},g\textsubscript{1})) at the same global $\overline{\varepsilon}_{xx}$. The simulation behavior is more diffuse and thereby fails to reproduce the exact experimentally observed localization band. {Several factors may be at the root of this. Primarily, the discreteness of plasticity is not captured well since continuum mechanics CP simulations generally do not include the discrete and stochastic nature of dislocation sources at the level of individual grains, unless introduced explicitly in the CP simulation framework (as was done in the work of Wijnen \textit{et al.}). Moreover, it is well known that it is hard to obtain a quantitative match between these experiments and simulations due to the unknown subsurface microstructure, see for example  Refs. \cite{TASAN2014386, Tasan2014, Diehl2017CrystalMicrostructure}. Even though the agreement between the strain fields is suboptimal, we do want to emphasize that (i) the global stress-strain curves match well (see \autoref{fig:fig1method}(e)), (ii) the majority of our strain fields show no strong mismatches between experiments and simulations and (iii) the activation of IHP/OHP plasticity, which is most important in our analysis and for the final conclusions, seems to be rather accurate.}

The simulations of ROI D2 do show a strain localization through the thinnest part of the notch, which is in reasonable qualitative agreement with experimental results (\autoref{fig:Paper_DROI2}(d)), considering the large amount of potential reasons for quantitative differences between experiments and CP simulations \cite{TASAN2014386, Tasan2014,Diehl2017CrystalMicrostructure}.
For both damaged ROIs, simulation results indicate dominant OHP slip activity regardless of the choice of model (\cps\ or \cpm). Along with the dominant OHP activity, a large hydrostatic stress concentration can be observed at the thin part of the notch (subplots (f\textsubscript{4}) and (g\textsubscript{4})), which has previously been discussed to be an indicator for damage \cite{TASAN2014386, hoefnagels2015retardation}. The lack of difference between the \cpm\ and \cps\ models is as expected: removal of a soft habit plane mechanism does not change the simulation results due to the low IHP BSF, which caused IHP activity to be very low also in the \cpm\ simulation. On balance, we conclude that an unfavorable habit plane in a martensite notch prevents enhanced martensite plasticity and thereby potentially increases the likelihood for damage.

\subsection{ROIs With Large Plasticity Without Damage}
\label{ROIplasticity}
\begin{figure}[H]
    \centering
    \includegraphics[width=\linewidth]{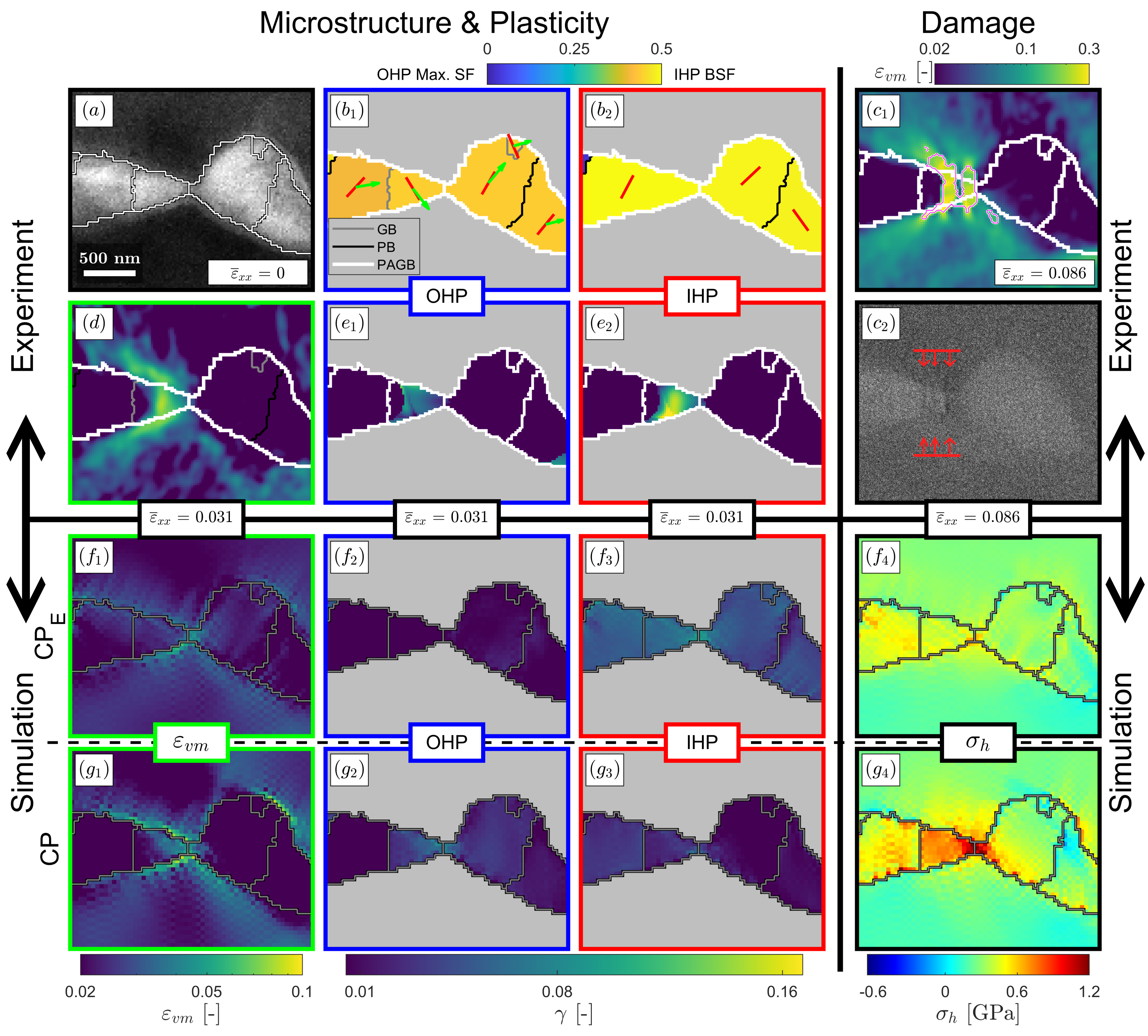}
    \caption{Overview of experimental-numerical results for ROI P1. Subplots are described in the caption of \autoref{fig:Paper_DROI1}. 
    (c\textsubscript{1}) Equivalent strain map for the final deformation increment, in which uncorrelated regions are indicated by the areas surrounded by pink boundaries; within these areas, the displacements were interpolated.
    The three sets of red arrows in (c\textsubscript{2}) point towards parallel, darker bands, which pass through the martensite into surrounding ferrite, and are ascribed to plastic localization bands.
    }
    \label{fig:Paper_PROI1}
\end{figure}

\begin{figure}[H]
    \centering
    \includegraphics[width=\linewidth]{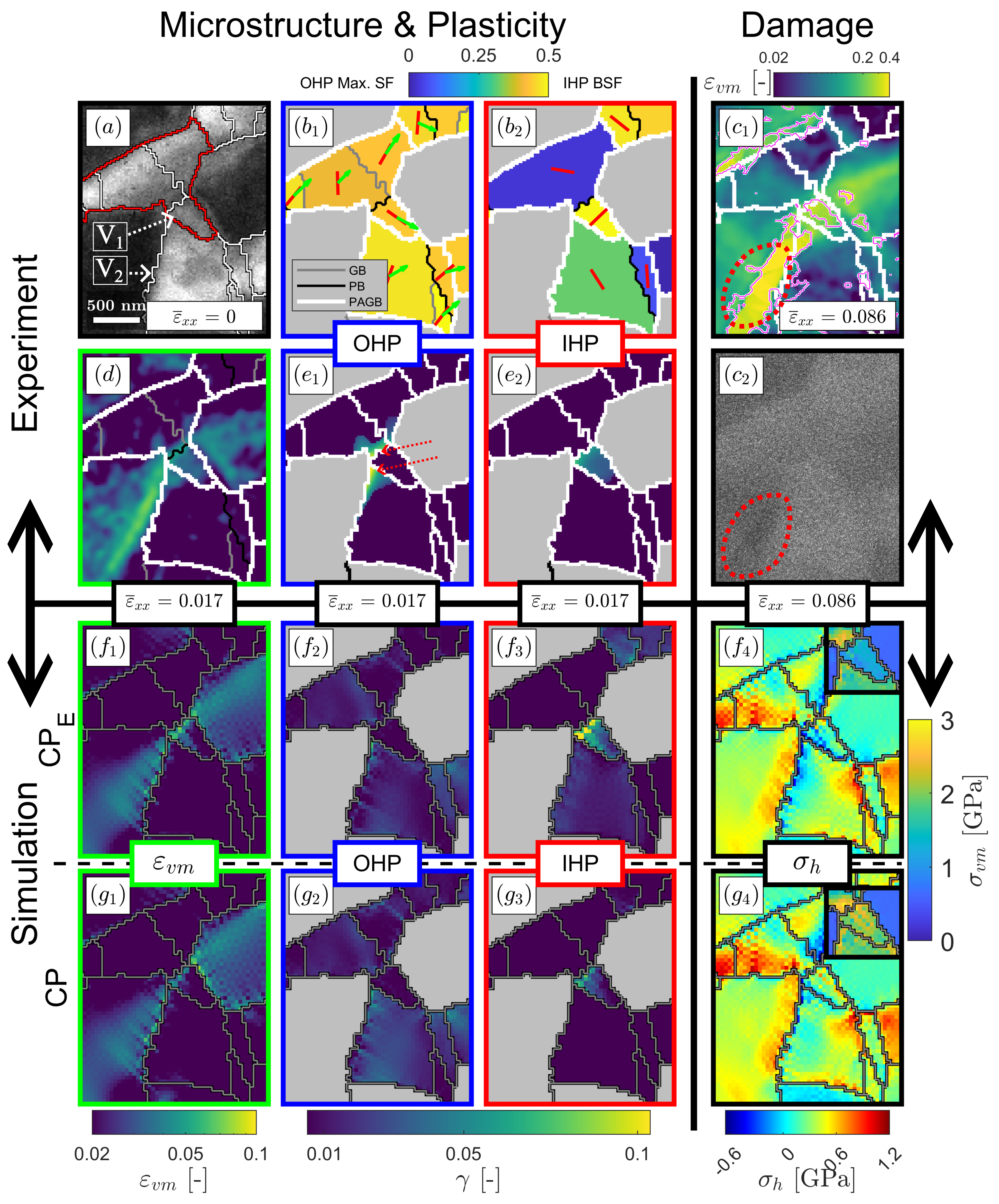}
    \caption{Overview of experimental-numerical results for ROI P2. Subplots are described in the caption of \autoref{fig:Paper_DROI1}. 
    (a) The red boundary indicates a PAGB for a PAG discussed in the text. Two relevant variants, V\textsubscript{1} and V\textsubscript{2}, are indicated by dashed white arrows. 
    (c\textsubscript{1}) Equivalent strain map for the final deformation increment, in which uncorrelated regions are indicated by the areas surrounded by pink boundaries; within these areas, the displacements were interpolated.
    (c\textsubscript{2}) A darker region in the in-lens BSE image is indicated by a dashed red circle, which is ascribed to large ferrite plasticity, at the location indicated in (c\textsubscript{1,2}).
    In (e\textsubscript{1}), the red dashed arrows indicate locations of comparatively strong OHP activity very near the boundaries of V\textsubscript{1}, but still in the adjacent martensite variants. Subplots (f\textsubscript{4},g\textsubscript{4}) include insets of the von Mises stress, $\sigma_{vm}$, around the martensite notch.}
    \label{fig:Paper_PROI2}
\end{figure}

\begin{figure}[H]
    \centering
    \includegraphics[width=\linewidth]{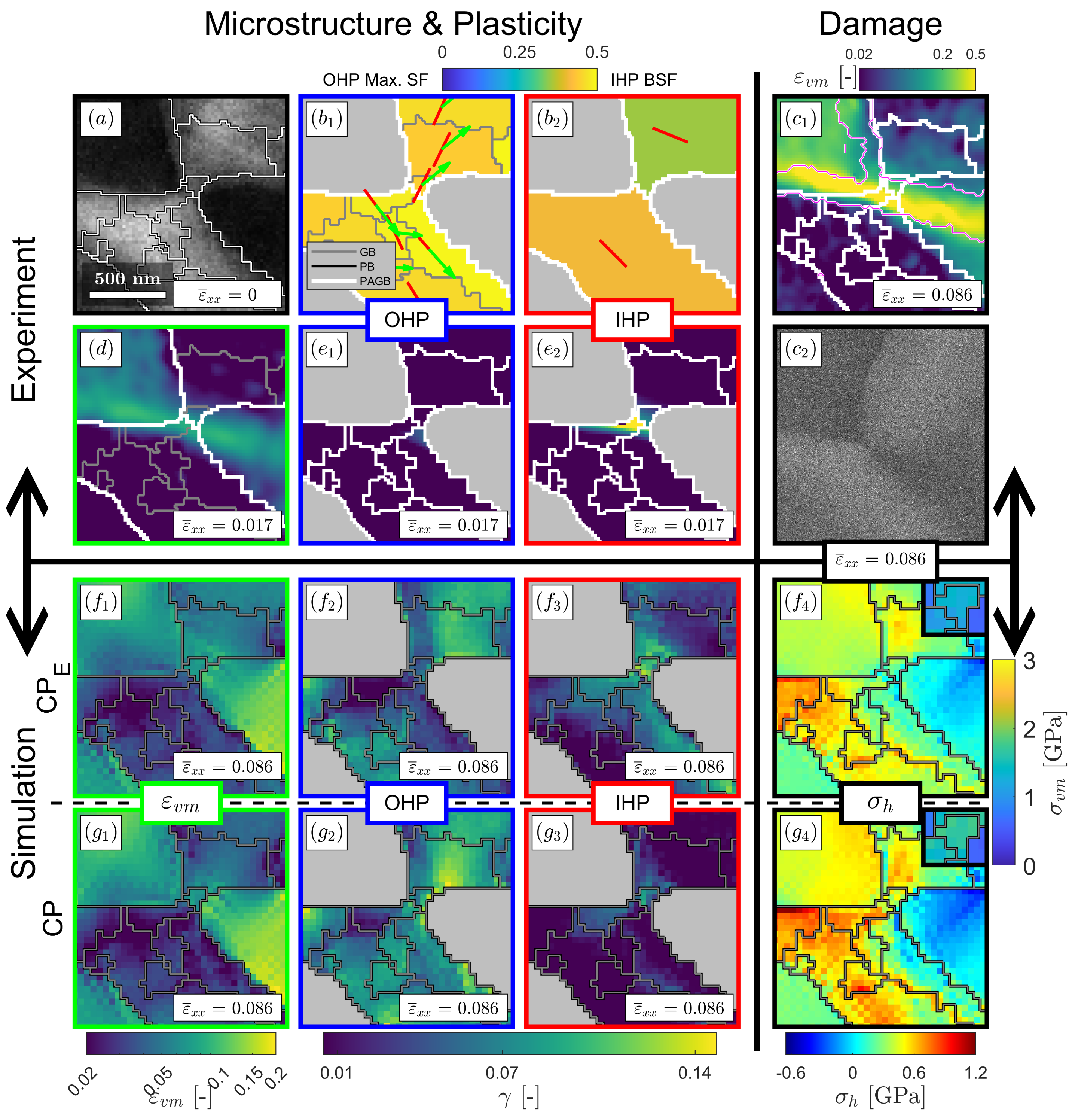}
    \caption{Overview of experimental-numerical results for ROI P3. Subplots are described in the caption of \autoref{fig:Paper_DROI1}. 
    (c\textsubscript{1}) Equivalent strain map for the final deformation increment, in which uncorrelated regions are indicated by the areas surrounded by pink boundaries; within these areas, the displacements were interpolated.
    (f\textsubscript{1,2,3}) and (g\textsubscript{1,2,3}) show simulation results for which the average strain  $\overline{\varepsilon}_{xx}$ closely matches that of the final experimental increment, as indicated in (f\textsubscript{1}), which means the simulation results do not correspond to the same deformation increment as displayed in (d).
    Subplots (f\textsubscript{4},g\textsubscript{4}) include insets with the von Mises strain, $\sigma_{vm}$, for the surroundings of the martensite notch.}
    \label{fig:Paper_PROI3}
\end{figure}

The three martensite notches that are potentially sensitive to damage, yet show significant plastic deformation without damage are displayed in \autoref{fig:Paper_PROI1} (ROI P1), \autoref{fig:Paper_PROI2} (ROI P2), and \autoref{fig:Paper_PROI3} (ROI P3). 

ROI P1 shows several similarities to ROI D1 \& D2, since all three ROIs contain a horizontally oriented notch with a PAGB on the thinnest part, indicating a high damage sensitivity. In contrast to ROI D1 \& D2, however, all variants in the notch are highly favorable for both OHP and IHP slip (\autoref{fig:Paper_PROI1}(b\textsubscript{1},b\textsubscript{2})). Moreover, in the in-situ BSE images, no damage is observed. The displayed deformation increment does show, similarly to ROI D1 \& D2, a plastic strain localization that does not pass through the thinnest part of the notch. \SI\ analysis reveals that the deformation is almost fully caused by IHP slip activity (\autoref{fig:Paper_PROI1}e\textsubscript{2}).

For ROI P1, the comparison between the \cps\ and \cpm\ simulations does reveal large differences. First, it is noticeable that the strain in the martensite variants in the \cpm\ model (\autoref{fig:Paper_PROI1}(f\textsubscript{1})) is more diffuse, compared to the \cps\ model (\autoref{fig:Paper_PROI1}(g\textsubscript{1})), in which most strain resides in the central two variants of the notch. {As argued in Section \ref{ROIdamage}, for case ROI D1, differences (in diffuseness) between experiments and simulations are not unexpected, since the subsurface microstructure and dislocation source stochastics are missing, yet the IHP/OHP plasticity ratio is sufficiently accurate for our purpose.}
The slip activity fields (\autoref{fig:Paper_PROI1}(f\us{2},f\us{3},g\us{2},g\us{3})) show that for \cpm\ simulations, IHP activity is dominant, whereas for \cps\ simulations, OHP and IHP activity both contribute (recall that IHP activity in the \cps\ model is fully carried by BCC crystallographic slip). As a result, hydrostatic stresses increase significantly in the \cps\ model (\autoref{fig:Paper_PROI1}(g\us{4})), to the same levels observed in the simulations of ROI D1 \& D2. In other words, the \cps\ model without a SBS mechanism predicts that damage will initiate in this martensite notch. Conversely, modelling an additional soft mechanism along the habit plane in the \cpm\ model significantly reduces hydrostatic stresses in ROI P1 (\autoref{fig:Paper_PROI1}(f\us{4})), thanks to the favorable habit plane orientation of all variants in the notch (\autoref{fig:Paper_PROI1}b\us{2}), which explains why damage does not occur in the experiment. This is direct evidence that the strong plastic deformation in experiments, which is carried by IHP activity as identified using \SI, has prevented damage initiation in the notch. 

Next, we consider the in-situ BSE image of the final increment of deformation. Several parallel, vertical, darker bands can be noted, marked by the red arrows in \autoref{fig:Paper_PROI1}(c\textsubscript{2}), on the left part of the notch. Since these bands (i) align with the large plastic strain band observed in the martensite (\autoref{fig:Paper_PROI1}(c\textsubscript{1})), and (ii) pass through the martensite into the surrounding ferrite (where they become more diffuse), the bands are not a result of damage initiation, but instead, of localized plastic deformation ($\sim31\%$ as indicated in \autoref{fig:eqStrains}(b)). Additionally, SF analysis has shown that the optimal slip direction along the (nearly vertically oriented) habit plane is pointed out of the sample plane, which means that the observed IHP activity is accompanied by out-of-sample-plane slip steps. This explains why the plastic localization bands are visible as darker bands in the BSE image.

The second notch that displays strong plasticity is displayed in \autoref{fig:Paper_PROI2} (the same notch was shown \autoref{fig:fig1method}, \ref{fig:PAGrecon} and \ref{fig:sslip}). Its orientation with respect to the direction of tension is $\sim 45-60 \degree$. As shown by the experimentally obtained strain map, a broad diffuse strain band passes through the notch, with maximum strains of $\sim37\%$, without resulting in observable damage. A slightly darkened region may be noted in the bottom left of \autoref{fig:Paper_PROI2}(c\textsubscript{2}) which coincides with a strong strain localization in the ferrite, indicated by the dashed red circles in \autoref{fig:Paper_PROI2}(c\textsubscript{1,2}). PAG reconstruction was challenging in this ROI: the PAG indicated by the red boundary in \autoref{fig:Paper_PROI2}(a) has an average misorientation of $5.8 \degree$ to the OR, however, this misorientation is still significantly smaller than that of all other PAG reconstruction possibilities. 
Given that variants at the edge of a PAG are reported to have a larger misorientation to the parent, this provides sufficient confidence that the PAG construction has been achieved successfully. 
SF analysis displays a high IHP BSF for the central variant in the notch (V\textsubscript{1}) but not for the surrounding packets \autoref{fig:Paper_PROI2}(b\us{2}), while all variants display a large OHP SF \autoref{fig:Paper_PROI2}(b\us{1}). \SI\ analysis confirms that IHP slip is active in the central grain, V\textsubscript{1}, while OHP slip occurs in the surrounding variants, most noticeably in the variant V\textsubscript{2} directly below V\textsubscript{1}, as has also been shown in detail in \autoref{fig:sslip}. However, due to the limited spatial resolution of the EBSD data, the exact locations of the grain boundaries between V\textsubscript{1} and its adjacent martensite variants has some uncertainty, meaning that the OHP activity of the two neighbouring variants, which peaks at the boundary region with variant V\textsubscript{1}, might also be IHP activity of V\textsubscript{1}. This also explains why the slip activity of the adjacent variants increases significantly close to the boundaries of V\textsubscript{1}, as indicated by the dashed red arrows in \autoref{fig:Paper_PROI2}(e\textsubscript{1}).

Simulations using the \cpm\ and \cps\ model show a strain localization in the notch, yet in both cases the localization consists of a thin band through the notch instead of more diffuse slip. Likely, plasticity initiates in the ferrite phase, and then localizes over the thinnest part of the notch, which is the point of direct contact between V\textsubscript{1} and the surrounding ferrite, which appears to be the path of least resistance. In contrast, the experimentally observed wider strain band may be caused by the unknown subsurface geometry. The slip activation for both \cpm\ and \cps\ simulations and the experiment compare well qualitatively, with the \cpm\ model predicting larger and more diffuse plasticity in V\textsubscript{1} because of the presence of a soft FCC mechanism. The, perhaps unexpected, noticeable localization in the \cps\ model variant V\textsubscript{1} likely occurs because the highest in-habit-plane SF for BCC slip is higher than the maximum out-of-habit-plane SF, even in the absence of a Substructure Boundary Sliding (SBS) mechanism. Nevertheless, the \cps\ model shows higher hydrostatic stresses compared to the \cpm\ model, though not at the same levels of ROI D1 \& D2. Again this is likely caused by the high SF of the in-habit-plane BCC slip systems and by of the orientation of the notch with regard to the direction of tension ($\sim 45-60 \degree$). A large difference is observed, however, for the equivalent von Mises stress, as is shown in the insets in \autoref{fig:Paper_PROI2}(f\textsubscript{4},g\textsubscript{4}), indicating that the SBS mechanism is able to release the stress level in the notch, thereby delaying damage initiation.

It is concluded that the martensite notch ROI P2 is able to deform strongly without damage initiation, which is attributed to a favorable habit plane within the central variant V\textsubscript{1}. Deformation of this central grain, enabled by its favorable habit plane orientation, is hypothesized to initiate strong OHP plasticity in the surrounding martensite variants. This idea is supported by the observation that the strain in V\textsubscript{1} occurs across the entire area of the variant, whereas surrounding variants such as V\textsubscript{2} only deform at the necessary locations, i.e. the locations where they share a grain boundary with V\textsubscript{1}. Overall, strains of $\sim37\%$ are reached without showing damage. \\

The third region, ROI P3 (\autoref{fig:Paper_PROI3}), also shows strong plastic deformation and the notch is also oriented at $\sim 45-60 \degree$ with respect to the direction of tension (i.e. maximum shear). It consists of a thin ($<100$ nm) meeting point of two PAGs with rather favorably oriented habit planes, yet the maximum OHP SFs are noticeably higher. The notch shows a very high strain of up to 70\% (\autoref{fig:eqStrains}(b), \autoref{fig:Paper_PROI3}(c\textsubscript{1})) without any sign of damage. The in-situ BSE image at the final increment of deformation reveals a darkened band at the center of the notch, which continues in the neighbouring ferrite grains, indicative of strong localized plastic deformation. \SI\ analysis reveals that, despite the significantly higher SF for OHP slip, the deformations are primarily carried by IHP activity.

Interestingly, the simulations at the same $\overline{\varepsilon}_{xx}$ as the first experimental increment of localization do not reveal significant strain localization in the relevant notch. Because of the way the strain paths formed in the simulation, strain did not concentrate in the ROI, despite the large simulated geometry of \arlarge\  and the use of isotropic buffer layer, as explained in \ref{sec:app_microstructure}. Possibly, substructure effects from experiments, which were not incorporated in the simulations, play a role as well. Simulation results at the final experimental increment ($\overline{\varepsilon}_{xx} = 0.086$), as shown in \autoref{fig:Paper_PROI3}(f\textsubscript{1,2,3}) and (g\textsubscript{1,2,3}), reveal some strain in the notch, which is more diffuse than the experimentally observed localization. Nevertheless, the comparison of the \cpm\ and \cps\ models for this simulation increment does yield clear results: in the \cpm\ model, the IHP activity is much more prominent than the amount of OHP activity in both the upper and lower PAG. Contrastingly, in the \cps\ model, the IHP activity is significantly lower than the OHP activity in both PAGs. Correspondingly, the stress maps in \autoref{fig:Paper_PROI3} (f\textsubscript{1}) and (g\textsubscript{1}) reveal that using the \cps\ model results in an increase in both von Mises stress and hydrostatic stress for both PAGs. Therefore, similar characteristics to the previously analyzed ROIs are observed in these simulations when comparing the two material models, even though the strain localization across the martensite is not as prominent due to the geometry of the simulated microstructure.

\subsection{Detailed Discussion of Slip Activity in Experiments and Simulations}

\noindent The apparent contrast between early martensite damage initiation and large plasticity has been analyzed in detail based on five ROIs. It was clearly observed that the habit plane orientation has a strong influence on the level of plastic deformation, though other noticeable observations were made as well. First, the largest strains are not always located at the thinnest part of the notch (clear examples are ROI D2 in \autoref{fig:Paper_DROI2} and ROI P1 in \autoref{fig:Paper_PROI1}). This suggests a possible resistance for plastic deformation along the PAGB, although the characteristics of strain localizations can also be influenced by the (unknown) subsurface morphology. Previous work suggests that martensite damage is more likely to occur along the Mn-embrittled PAGB \cite{archie2017micro, Kuzmina2015GrainSteel}. However, all damaged and non-damaged notches in this analysis contain a PAGB (that is situated at the thinnest part of the notch), which does not appear to have an influence on damage initiation. In fact, the strain band shown for ROI P2 in \autoref{fig:Paper_PROI2} passes perpendicularly through a PAGB, while the strain band of ROI P3 in \autoref{fig:Paper_PROI3} passes on both sides of the PAGB, oriented parallel to the strain band, both without showing any sign of damage. Note that the high spatial resolution of the SEM-DIC was crucial in being able to distinguish the location of the strain bands, just next to the PAGB. In all, it seems that the prior austenite grain boundary is not as damage-sensitive as previously presumed, and that other factors such as notch geometry and habit plane orientation are more relevant for damage initiation.\\

In regions such as ROI P3, slip along the habit plane was observed despite an apparent lower favorability, when compared to OHP slip systems. This is in line with findings of Du \textit{et al.}, who stated that SBS may occur at seemingly unfavorable orientations \cite{Du2019}. However, their findings were obtained for nano-tensile tests of single packets isolated from fully martensitic steel and have only been indirectly shown to hold for DP steel microstructures \cite{Du2019,Du2018a}. Activation of particular martensite slip systems in DP steel may be constrained by activation of slip systems in ferrite, through which plasticity in DP is generally thought to initiate \cite{Tasan2014}. Despite these slip compatibility and/or geometric constraints, it is demonstrated in this work that activity along the habit plane is preferred, even at unfavorable orientations. This also shows that the activation or inactivation of slip in martensite seems to be the governing factor on where a larger deformation band over a martensite notch can form.

\begin{figure}[H]
    \centering
    \includegraphics[width=0.9\linewidth]{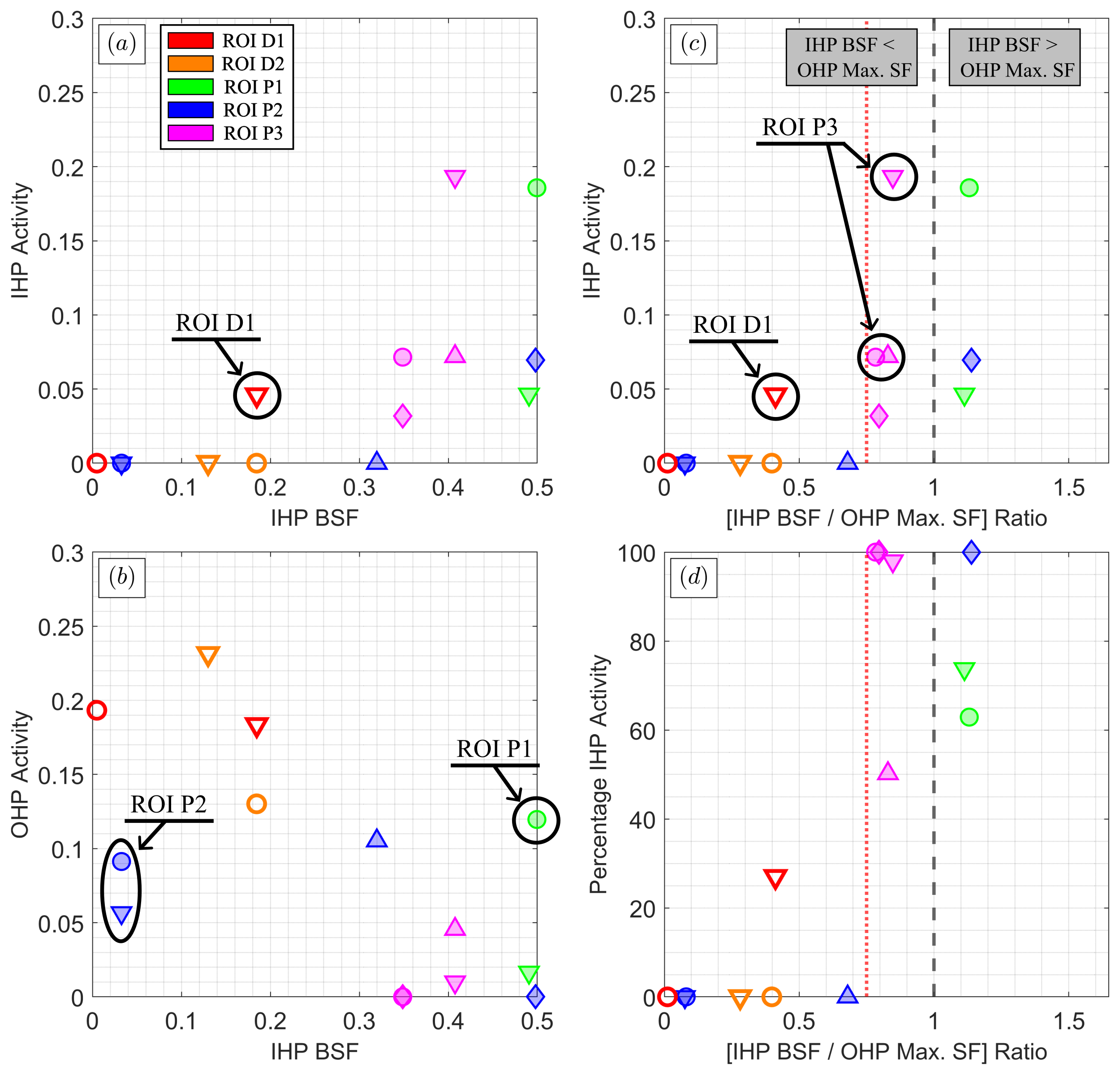}
    \caption{Correlations between maximum out-of-habit-plane Schmid Factor (Max. OHP SF) and in-habit-plane Boundary Sliding Favorability (IHP BSF) values, and slip activation for variants in martensite notches, displayed for the first increment of noticeable deformation. (a) IHP activity as a function of BSF. (b) OHP activity as a function of BSF. (c) IHP activity as a function of the ratio IHP BSF / Max. OHP SF, where the dashed line indicates a ratio of 1 and the red dotted line indicates a ratio of $0.75$. (d) Percentage of IHP activity of the total slip activity (OHP+IHP) in each variant, plotted against the ratio IHP BSF / Max. OHP SF; a sharp transition is observed at a ratio of $\sim 0.75$ (see red dotted line). Red and orange symbols indicate variants from ROIs with damage; blue, purple and green symbols represent ROIs without damage (see the legend). The shape of the icons indicates the variants in each ROI, see \autoref{fig:eqStrains}. Encircled data points highlighted by arrows are discussed in the text.
    }
    \label{fig:expDisc}
\end{figure}

To further analyze the relation between habit plane favorability and IHP slip activity, the results of the \SI\ analysis for the five notches discussed in \autoref{ROIdamage} and \autoref{ROIplasticity} are gathered and shown in \autoref{fig:expDisc}. The displayed slip activities in this plot indicate the maximum activity in each variant, calculated as the mean of the 15 pixels with largest slip magnitude. The two damaged regions are indicated by the colors red and orange, and the three non-damaged regions are colored blue, purple and green, whereas the symbols represent the different variant of each ROI, similar to \autoref{fig:eqStrains}. First, \autoref{fig:expDisc}(a) shows a clear separation in the graph with all damaged variants on the left (low IHP BSF) and all but two non-damaged variants on the right (high IHP BSF), suggesting that damage does not initiate when the BSF is high for one of the variants in the notch, i.e., it only occurs when all variants in the notch have a low BSF. Secondly, the graph shows a positive correlation between BSF and IHP activity, which is to be expected because a higher habit plane favorability increases the resolved shear stress on the habit plane. Non-damaged regions show much higher IHP activity compared to damaged regions, with only limited IHP activity for ROI D1 at a BSF of $\sim 0.2$. 

\autoref{fig:expDisc}(b) shows a distinctive trend as well: a higher IHP BSF results in a lower OHP activity. The separation between damaged and non-damaged regions in levels of OHP activity is apparent: the damaged regions have a higher OHP activity than the non-damaged regions. There is a non-damaged variant from ROI P1, as indicated in \autoref{fig:expDisc}(b), which shows strong OHP activity (along with high IHP activity) despite a high IHP favorability. Since this notch consists of two packets, this high OHP activity in one of the packets is likely caused by the fact that the other packets deforms strongly along the habit plane. \autoref{fig:expDisc}(c) shows the ratio between IHP BSF and maximum OHP SF (referred to as the IHP/OHP ratio). A ratio lower than 1 indicates a seemingly unfavorable habit plane orientation. Nevertheless, IHP activity is still observed for ratios below 1, especially for the variants of ROI P3, whose IHP/OHP ratio is $\sim 0.75$, at which we plot a red dotted line in \autoref{fig:expDisc}(c,d). This behavior is similar to what was reported by Du \textit{et al.} \cite{Du2019}, showing that IHP slip activates more easily than OHP slip, at seemingly unfavorable orientations. The IHP activity for ROI D1, as indicated by the arrows in \autoref{fig:expDisc}(a) and (c), is noteworthy for its low IHP/OHP ratio. However, the IHP activity is not very large and also outside the thinnest notch section, therefore, it is not surprising that this limited IHP activity is unable to prevent damage initiation in the notch center, where the stresses are much higher. Nonetheless, this IHP activity does show that habit plane activation is possible at IHP/OHP ratios far below those reported in previous literature. Finally, \autoref{fig:expDisc}(d) shows the percentage IHP activity of the total sum of slip activity in each variant, plotted against the IHP/OHP ratio. This measure of IHP slip activity is less sensitive to small strain concentrations and also independent of the total strain in the variant, which allows for better comparison between the ROIs. The observed trend from \autoref{fig:expDisc}(a) and (c) becomes sharper in (d): an IHP/OHP ratio above $\sim 0.75$ (see the red dotted line) results in dominant IHP activity, whereas a ratio below $\sim 0.75$ results in dominant OHP activity. For damaged cases, OHP activity is dominant, while for most of the variants in the non-damaged cases, IHP activity is dominant.

\begin{figure}[H]
    \centering
    \includegraphics[width=0.8\linewidth]{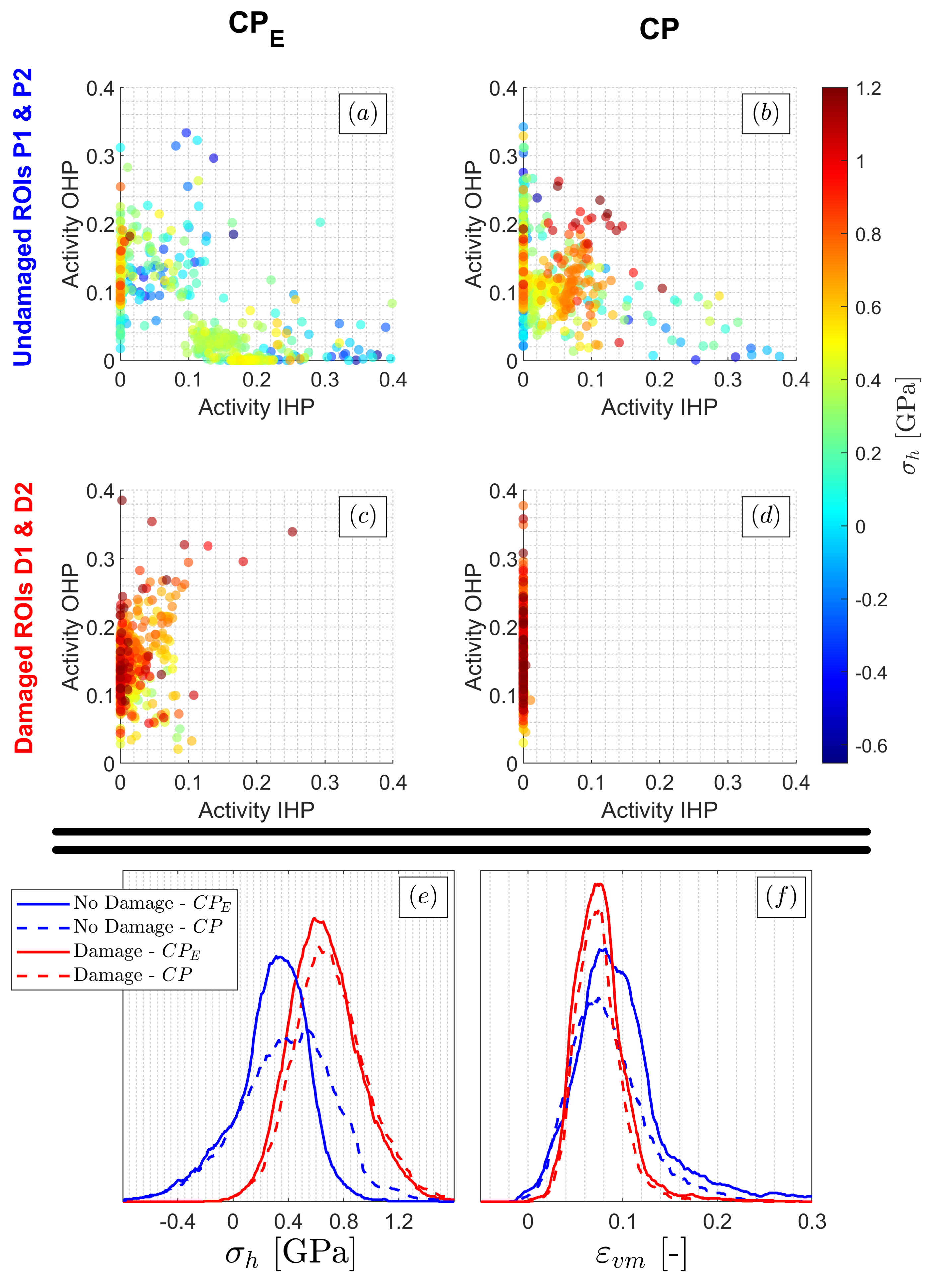}
    \caption{Comparison of hydrostatic stress (color scale) versus IHP/OHP activity for the \cpm\ (a,c) and \cps\ models (b,d). All data are taken from the simulation increment whose average $\overline{\varepsilon}_{xx}$ most closely matches the final experimental increment, $\overline{\varepsilon}_{xx} = 0.086$. The IHP activity is plotted against OHP activity for each martensite simulation voxel that is within 500 nm of the center of each martensite notch. The color of each data point indicates the hydrostatic stress level, corresponding to the colorbar. (a,b) Data points for the notches that did not damage, (c,d) data points for notches that damaged. (e,f) Distribution plot of the (e) hydrostatic stress and the (f) equivalent strain for \cpm\ (full lines) and \cps\ (dashed lines) models for the non-damaged (blue lines) and damaged notches (red lines). }
    \label{fig:numDisc}
\end{figure}

Habit plane favorability and the possible activation of SBS are hypothesized to have an influence on (hydrostatic) stress levels, and thereby on damage initiation and subsequent damage evolution. We analyze this by performing a statistical comparison between the \cpm\ and \cps\ models. The effect of modelling substructure boundary sliding along the habit plane may be uncovered by analyzing both IHP and OHP activity, and the corresponding levels of hydrostatic stress and equivalent strain, in the two models. This analysis is shown in \autoref{fig:numDisc}, in which (a) and (b) contain data for non-damaged ROIs P1 \& P2, while ROI P3 was not included due to the poor experimental-numerical match. Each data point represents a martensite voxel within a radius of 500 nm of the center of the martensite notch, and is colored to reflect the hydrostatic stress state. The same data for the damaged ROIs, D1 \& D2, is displayed in \autoref{fig:numDisc}(d,e). Comparison of \autoref{fig:numDisc}(a) and (b) shows that using a \cpm\ model results in more IHP activity, at lower hydrostatic stress levels. Note that the \cpm\ simulation still underpredicted the IHP slip magnitudes observed in experiments, therefore, the real hydrostatic stress in the undamaged notches is probably significantly lower than the simulated values in \autoref{fig:numDisc}(a). Additionally, in the \cpm\ model, voxels with dominant IHP activation show on average a lower stress level than voxels with dominant OHP activation. Using the \cps\ model, a substantial increase in hydrostatic stress, to unrealistic levels, is noted for the non-damaged regions, which is paired with a decrease in IHP activity. When switching from the \cpm\ to the \cps\ model, the hydrostatic stress distribution in \autoref{fig:numDisc}(e) shows an increase in stress, while the strain distribution in \autoref{fig:numDisc}(f) shows a decrease in magnitudes. The likelihood of damage is increased as concluded from (i) the general increase of hydrostatic stresses to unrealistic levels and (ii) the increase in stress heterogeneity. Note that the hydrostatic stress distribution for ROI P1 \& P2 using the \cps\ model (blue dashed line in \autoref{fig:numDisc}(e)) reaches similar maximum values as observed for ROI D1 \& D2 with the \cpm\ model (full red line in \autoref{fig:numDisc}(e)). This further supports the hypothesis that the \cps\ model incorrectly increases the probability for damage in all martensite notches.

For the damaged regions ROI D1 \& D2, the \cpm\ model predicts high hydrostatic stresses but allows for some level of IHP activity. The \cps\ model does not show significant IHP activity, which corresponds with a small increase in hydrostatic stress, as observed in \autoref{fig:numDisc}(e). The strain distributions for the two models are similar, showing only a slight decrease when using the \cps\ model. The lack of a strong difference in strain distributions between the models for damaged ROIs is expected, since IHP activity was already low in the \cpm\ model in ROI D1 \& D2. In general, the \cpm\ approach predicts lower levels of hydrostatic stress at larger levels of IHP activity, which is most notable for the non-damaged ROIs P1 \& P2. However, removing the SBS mechanism by using the \cps\ model yields similar stress states for undamaged ROI P1 \& P2 and damaged ROI D1 \& D2. Since damage initiation does not happen in the experiment in ROI P1 \& P2, the increased stress state in the \cps\ model provides a strong indication that there is indeed a soft habit plane mechanism active in the real DP600, which allows for large IHP activity to occur and stresses to remain comparatively low. On balance, this suggests that a favorable habit plane orientation in martensite notches in DP600, resulting in stress release by means of substructure boundary sliding, is an important damage inhibitor. 

\newpage
\section{Conclusions}

Damage initiation in the lath martensite phase of DP steel has been studied extensively in the literature. Yet, so far, the deformation mechanisms that lead to damage initiation remained mostly unexplored. Additionally, the apparent contrast between "brittle" martensite damage and large martensite plasticity (potentially related to the habit plane orientation and to substructure boundary sliding) had not yet been investigated. In this paper, a one-to-one experimental-numerical coupled approach was employed, using a series of novel experimental, analysis and simulation methods, to unravel the competition between large plasticity and damage initiation in martensite notches in DP steel. Specifically, high-resolution SEM-DIC, in combination with prior austenite grain reconstruction, Schmid factor analysis and a recently developed novel slip system identification method (SSLIP), was used to reveal active slip systems on the level of individual martensite variants, while monitoring the occurrence of damage. Numerical simulations of the experimental microstructures were performed using both a novel substructure \cpm\ approach and a \cps\ model to analyze the effect of a soft boundary sliding mechanism along the habit plane of martensite variants, specifically in relation to martensite damage. Analysis on 5 regions of interest revealed the following insights:

\begin{itemize}
    \item Several 'damage-sensitive' martensite notches are able to deform strongly ($30-70 \%$ local equivalent strain) without showing clear damage. 
    \item The large plasticity in martensite notches is shown to be enabled by in-habit-plane slip along the habit plane in the variants inside the notch. This behavior can be replicated in CP simulations by using the \cpm\ approach that incorporates the soft mechanism in lath martensite.
    \item Martensite notches in which damage initiates are shown to consist solely of variants with an unfavorable habit plane orientation, and can therefore only deform up to maximum $10 \%$ local strain by out-of-habit-plane slip activity before damage initiates.
    \item Habit plane slip activity in DP steel is shown to occur at seemingly unfavorable orientations with a sharp transition to dominant activation occuring at an in-habit-plane over out-of-habit-plane Schmid factor ratio of $> 0.75$, with even some in-habit-plane activity observed at a ratio of $\sim 0.4$. 
    \item All analyzed martensite notches contained a prior austenite grain boundary on the thinnest point, which appears to have no influence on damage initiation or the lack thereof, in clear contrast to the hypothesized central role of such boundaries on damage initiation in lath martensite in the literature.
    \item Numerical analysis, in which an enriched and a standard CP model were employed and compared, shows that addition of a soft mechanism along the habit plane is required to properly capture the behavior of the non-damaged martensite notches. Notably, in-habit-plane activity is required to provide release of the hydrostatic stress to levels below that of damage initiation.
    \item The wide range of consistent observations and analyses in this paper provides indirect but strong evidence for the premise that substructure boundary sliding is active in DP steel.
\end{itemize}

In summary, this contribution shows that activation or inactivity of habit plane plasticity is the main cause for the anisotropic behavior of martensite as observed in DP steel. A favorable habit plane orientation is a key factor in the inhibition of microscopic damage initiation in martensite notches which are conventionally sensitive to damage, a property which can be accurately captured in an enriched numerical model. 

The challenge now is to exploit the boundary sliding in the design and optimization of (new) multi-phase steels. {For example, tuning the martensite notches to consist of single packets will help to prolong SBS in cases where its habit plane is favorably oriented for the local stress state. In such a configuration, the activated SBS mechanism will not be blocked by surrounding harder packets and can deform plastically along with the surrounding ferrite. Alternatively, we recommend to optimize the overall prior austenite grain texture to promote habit plane aligned plasticity. This likely requires an approach in which calibrated Enriched CP simulations are performed on larger areas with many notches, to statistically quantify the propensity for martensite damage initiation for certain textures. Subsequently, this requires verification by statistical analyses of damage in experiments \cite{kusche2019, tang2021mesoscopic}. Thereby, new opportunities could arise for designing more damage-resistant steel grades.}

\section*{Author Contributions (CRediT)}
\textbf{Tijmen Vermeij:} Conceptualization, Methodology, Software, Investigation, Writing - Original Draft, Visualization, Supervision

\textbf{Casper Mornout:} Methodology, Software, Investigation, Writing - Original Draft, Visualization

\textbf{Vahid Rezazadeh:} Methodology, Software, Writing - Review \& Editing, Supervision

\textbf{Johan Hoefnagels:} Conceptualization, Methodology, Resources, Writing - Review \& Editing, Supervision, Funding Acquisition

\section*{Acknowledgements}
The authors acknowledge Niels van de Straat, Job Wijnen, Roy Kerkhof, Ron Peerlings, Marc Geers and Marc van Maris for discussions and experimental support.  

This research was carried out as part of the "UNFAIL" project, under project number S17012b in the framework of the Partnership Program of the Materials innovation institute M2i (www.m2i.nl) and the Netherlands Organization for Scientific Research (http://www.nwo.nl).

\appendix
\section{Detailed experimental and numerical methodology}
\label{App1}

\subsection{Grain reconstruction and extension}
\label{sec:App_recon}
Grains are reconstructed with the MTEX toolbox in MATLAB \cite{bachmann2010, MTEX}, with a misorientation threshold of 2.5$\degree$. The M-F phase distribution is obtained by thresholding CI data, with occasional manual selection of grains based on aligned ECCI/CI data. Grain boundary positions calculated from EBSD data show a clear mismatch to the phase information from ECCI/CI data on M-F boundary regions, with the ECCI/CI data systematically showing larger martensite islands, as seen in \autoref{fig:PAGrecon}(a\us{1}). This effect is particularly notable in vertical direction. Since ECCI and CI data more accurately show the M-F interface locations, an average of the two is calculated and smoothed with a moving average filter of 5 pixels. A threshold is applied on this ECCI/CI data to allow extension of martensite islands in vertical and horizontal direction, in an iterative procedure to ensure natural isotropic extension of grains. This grain extension step is vital for proper characterization of the shape of martensite notches, as shown in \autoref{fig:PAGrecon}(a\us{1}).

\subsection{Prior Austenite Grain Reconstruction}
\label{sec:App_pag}
PAG reconstruction is required for determination of the habit plane orientation of each variant. Prior to performing the actual PAG reconstruction, the best-fitting Orientation Relationship (OR) is found by performing the optimization routine of Nyyssönen \textit{et al.} \cite{Nyyssonen2016IterativeMisorientations} (included in MTex). Starting with an initial guess of KS OR, all available martensite variant-to-variant misorientations are used to optimize the OR, which is between KS and NW. Next, this OR is employed to perform PAG reconstruction using the recently introduced Variant Graph method \cite{hielscher2022variant, niessen2022parent}, a powerful reconstruction approach that considers all 24 possible parent orientations for each variant, whilst also performing next-neighbour analysis (i.e., not only considering direct neighbouring variants for the reconstruction), while remaining computationally efficient.

Despite the versatility of the Variant Graph method, PAG reconstruction in low martensite vol\% DP grades is challenging. Since the martensite islands are small, they usually have a low number of variants, rendering some resulting PAG orientations non-unique. Additionally, variants neighbouring a PAG boundary have been observed to potentially have a double KS-OR, sharing an OR with both parent orientations they neighbor \cite{Archie2018OnImplications}. Due to these difficulties, the Variant Graph approach has been observed to provide incomplete reconstructions in the material used in this study. This can be checked for one prior austenite grain (PAG\textsubscript{1}) by randomly generating a large number of uniformly distributed parent orientations PAG\textsubscript{R} with 1$^\circ$ misorientation between neighbouring PAG\us{R}. The 'fit' of each PAG\textsubscript{R} is defined as the average misorientation of each of the variants in PAG\textsubscript{1} towards the best-fitting theoretical child orientations of PAG\textsubscript{R}. If all the good-fitting PAG\textsubscript{R} orientations are clustered around one best-fitting PAG\us{R}, the orientation of PAG\textsubscript{1} is considered to be unique. If the good-fitting PAG\us{R} are grouped in multiple clusters, but they do share one matching $\{111\}_{\gamma}$ plane that is aligned with a $\{110\}_{\alpha'}$ plane of the variants within PAG\textsubscript{1}, then this $\{111\}_{\gamma}$ plane is still accepted as the habit plane for PAG\textsubscript{1}. All relevant PAGs were thereby manually checked and corrected where necessary, by moving individual variants across PAGs.

\begin{figure}[H]
    \centering
    \includegraphics[width=0.6\linewidth]{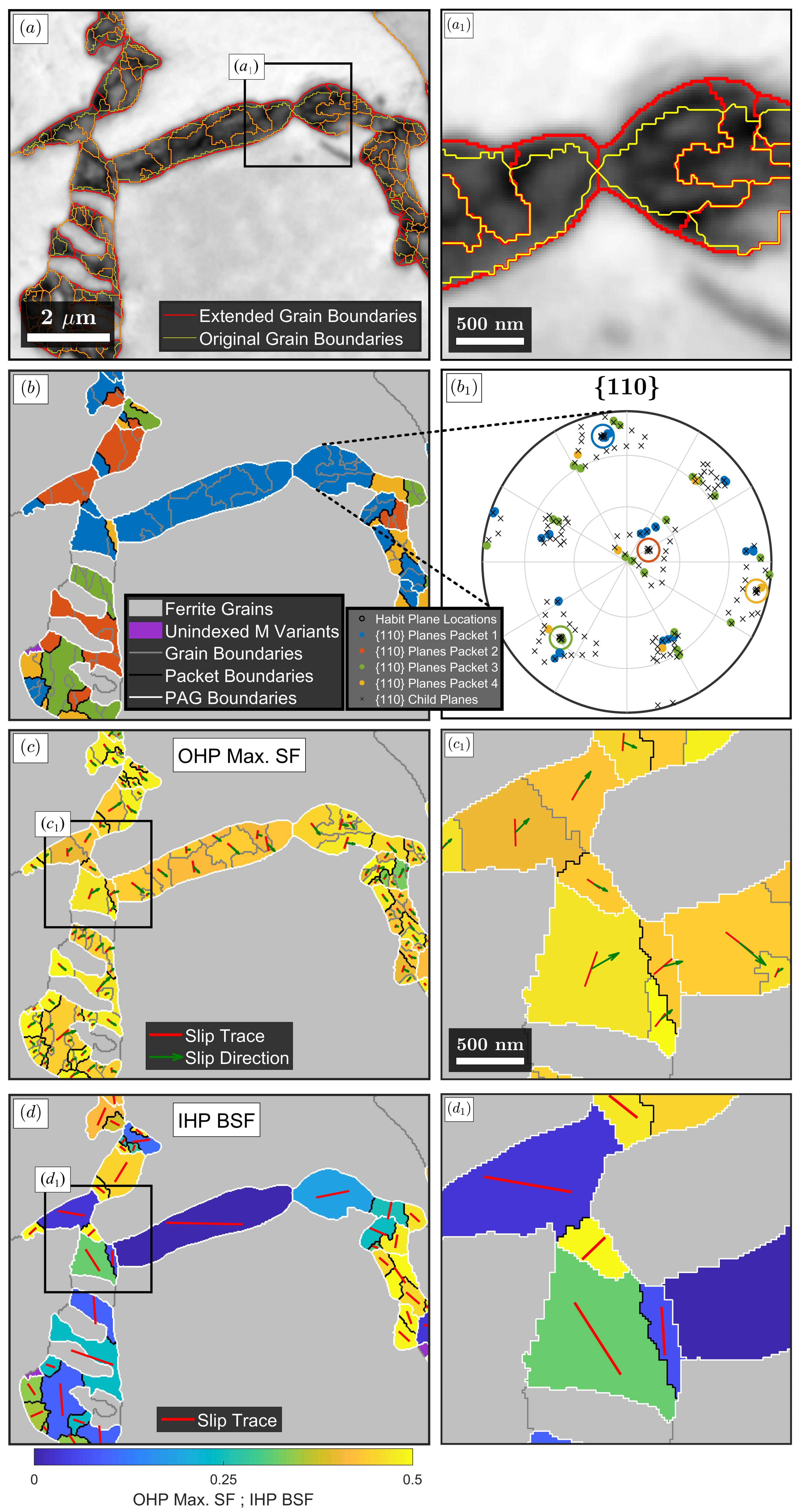}
    \caption{Overview of experimental analysis techniques applied on a single region. (a) Averaged and smoothed ECCI/CI data field, with original grain boundaries from EBSD in yellow and extended grain boundaries in red. (a\textsubscript{1}) Grain extension around a martensite notch. (b) Results of Prior Austenite Grain (PAG) reconstruction, where packets are given different colors. (b\textsubscript{1}) $\{110\}$ Pole figure of a PAG, in which the $\{110\}$ plane normal vectors of the experimental variants (filled colored circles) and the 24 theoretical children (black crosses) are indicated, as well as the locations of the habit planes (open colored circles). The dashed lines indicate from which PAG the pole figure originates; the colors in the pole figure correspond to the packet colors in this PAG. (c) Max. out-of-habit-plane (OHP) Schmid Factor (SF) map, including the 2D slip trace and slip direction for each variant. (d) In-habit-plane (IHP) boundary sliding favorability (BSF) for each packet, including 2D habit plane trace. (c\us{1}) and (d\us{1}) show an inset of a relevant notch.
    }
    \label{fig:PAGrecon}
\end{figure}

The PAG reconstruction for the region displayed in \autoref{fig:fig1method} is shown in \autoref{fig:PAGrecon}(b), in which PAGBs and packet boundaries are indicated. A $\{110\}$ pole figure for one PAG is included in (b\textsubscript{1}) which shows $\{110\}_{\alpha'}$ plane normal vectors for each of the experimental variants, as well as for the 24 theoretical children, displaying the quality of the fit and the locations of the habit planes (highlighted in \autoref{fig:PAGrecon}(b\textsubscript{1}) by large colored circles).

\autoref{fig:PAGrecon}(c,d) shows the magnitude of the Max. OHP SF and the BSF for IHP slip (as explained in Section \ref{sec:IdPlasticity}), respectively, including corresponding slip traces (and slip directions for OHP slip). The insets in (c\us{2},d\us{2}) give a more detailed view of these values.

\subsection{Slip System Identification (SSLIP)}
\label{sec:App_sslip}

\begin{figure}[H]
    \centering
    \includegraphics[width=\linewidth]{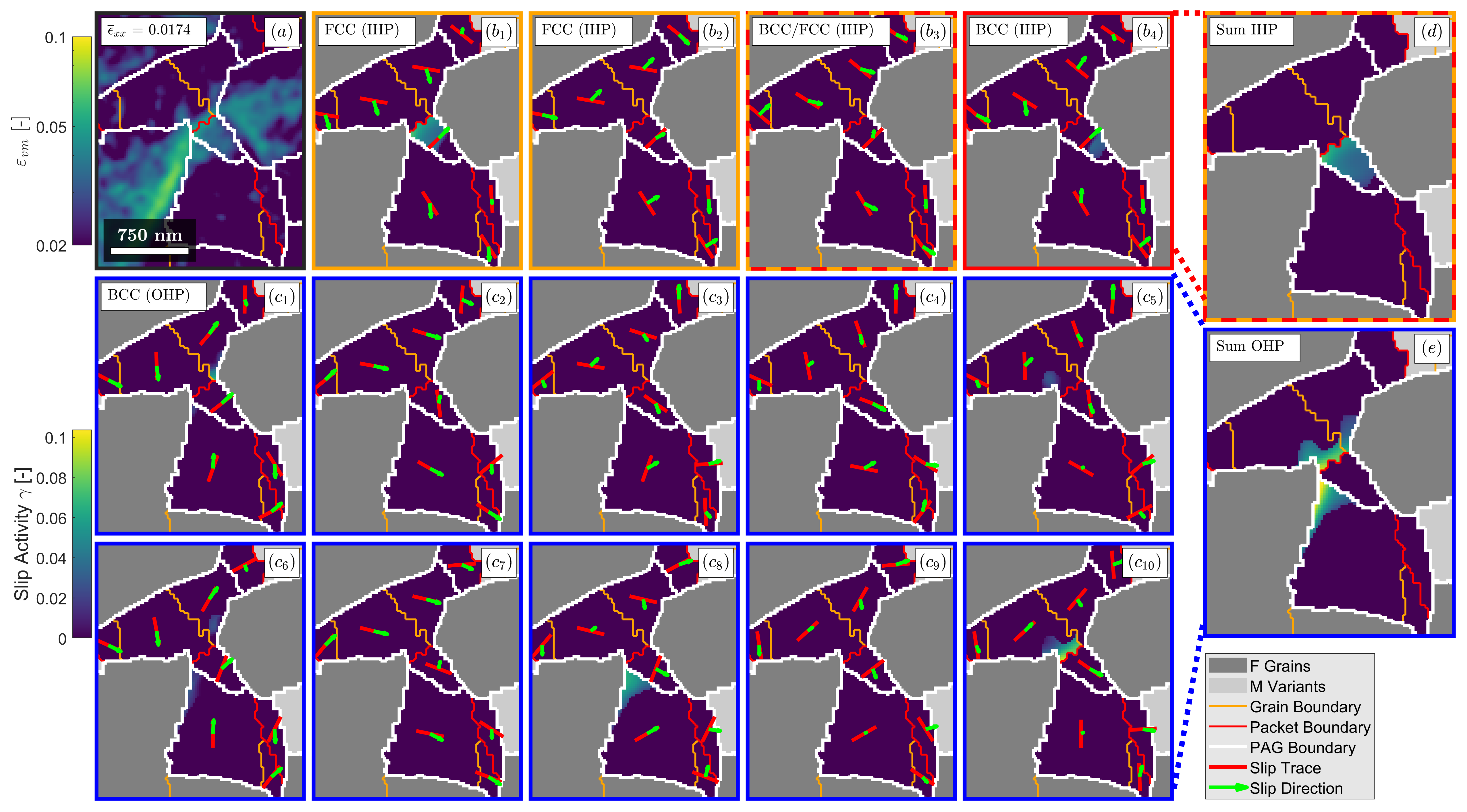}
    \caption{Overview of \SI\ identification performed individually on the variants in the ROI labeled \texttt{P} in \autoref{fig:fig1method}. Slip system activity fields are plotted together for all variants and grouped per slip type, with (b\textsubscript{1-4}) FCC IHP, BCC IHP and the matching BCC-FCC slip systems marked in red, orange and orange-red frames, and (c\textsubscript{1-10}) the 10 BCC OHP slip systems marked in blue frames. For each slip system in each variant, the 2D slip trace (red) and in-plane slip direction (green) are drawn. All slip activity maps are plotted on the same scale. (a) Equivalent strain map for the corresponding deformation increment, in which the global $\overline{\varepsilon}_{xx}$ is indicated. (d,e) The sum of IHP and OHP slip activity respectively, obtained by summing the subplots on the left with similarly colored frames.}
    \label{fig:sslip}
\end{figure}

For the identification of slip system activities, we employ a recently proposed, novel methodology: Slip System based Local Identification of Plasticity (\SI) \cite{vermeij2022crystallographic}. \SI\ identification uses the measured 2D displacement gradient tensor $\mathbf{H}^{exp}$ to estimate the most likely (combination of) active slip systems for each individual data point in the deformation map. It functions by matching $\mathbf{H}^{exp}$ to a theoretical displacement gradient tensor $\mathbf{H}^{theor}$, which is the combination of contributions of all potential slip systems $\alpha$, and is given by:
\begin{equation}
\mathbf{H}^{theor}=\sum_{\alpha=1}^{n} \mathbf{H}^{\alpha} = \sum_{\alpha=1}^{n} \gamma^{\alpha} \Vec{s}^{\:\alpha} \otimes \Vec{n}^{\alpha},
\end{equation}
where $\Vec{s}^{\:\alpha}$ and $\Vec{n}^{\alpha}$ are the slip direction and the normal vector of the slip plane of slip system $\alpha$. $\gamma^{\alpha}$ is the to-be-identified slip amplitude on slip system $\alpha$ and $n$ is the number of considered slip systems. The \SI\ method aims to describe the measured 2D deformation at each datapoint with a minimal amount of total slip, whilst keeping the difference between $\mathbf{H}^{exp}$ and $\mathbf{H}^{theor}$ below a certain threshold $H_{thresh}$, resulting in the following optimization problem:
\begin{mini!}|1|
{\gamma^{1},\ldots,\gamma^{n}}{\sum_{\alpha=1}^{n}{|\gamma^{\alpha}|},\label{eq:objfun}}
{\label{eq:optim}}{}
\addConstraint{||\mathbf{H}^{exp}-\mathbf{H}^{theor}||^{2D}}{<H_{thresh.}\label{eq:con1}}{}
\end{mini!}
The value of $H_{thresh}$ is individually optimized for each datapoint in the ROI in this work by choosing the lowest possible value at which a complete solution is found that is uninfluenced by noise in the DIC data. Note that the local point-by-point analysis approach removes the need for slip trace analysis. {Additionally, one could argue that the CRSS of the slip systems (which differ between austenite and martensite) could be added to the identification to minimize the total amount of work, instead of the total amount of deformation, as also suggested in Ref. \cite{vermeij2022crystallographic}, even though this ignores that ultimately the local presence of experimentally observed slip is determined primarily by the presence of a dislocation source, not the CRSS. We explored this option (by assuming a CRSS for austenite FCC that is 0.5 that of martensite BCC) and it did not appear to influence the results significantly. Therefore, it was decided to leave out a possible CRSS influence and let the slip system identification be purely based on the experimental observations.}
 In this work, the \SI\ method is employed to distinguish between in-habit-plane (IHP) and out-of-habit-plane (OHP) activity for each martensite variant in the analyzed ROIs. The \SI\ framework takes into account the 12 BCC $\{110\}\langle111\rangle_{\alpha'}$ slip systems, as well as 3 FCC $\{111\}\langle110\rangle_{\gamma}$ slip systems to represent the hypothesized slip along retained austenite films. However, since one $\{111\}\langle110\rangle_{\gamma}$ system very closely matches one of the BCC systems (due to the OR), the activity of these two overlapping slip systems cannot be distinguished, therefore, the matching $\{111\}\langle110\rangle_{\gamma}$ slip system is removed from the analysis to ensure convergence. Only datapoints with a 2D equivalent strain above 2\% are analysed. For the ROI labeled \texttt{P} in \autoref{fig:fig1method}(c\textsubscript{1,2}), an example of \SI\ identification is shown in \autoref{fig:sslip}. The 2D equivalent strain field of the used DIC increment is displayed as reference in \autoref{fig:sslip}(a). Note that \SI\ has been performed for each variant individually, due to their different orientation. However, the results are plotted together, and results for each variant have been sorted such that IHP and OHP systems for different variants are shown in the same subplots. For IHP, four slip systems are shown: 2 FCC slip systems (b\textsubscript{1,2}), one BCC slip system (b\textsubscript{4}) and the BCC/FCC matching slip system (b\textsubscript{3}). All OHP fields are plotted in no particular order in (c\textsubscript{1-10}). The total slip activity from both types, IHP and OHP, are summed per datapoint for the full map, the result of which is shown in \autoref{fig:sslip}(d,e). 

In this example, a diffuse strain band which passes through four variants, from two different PAGs, is assigned to different slip systems. The automatic \SI\ identification seems reliable, based on the following physics-informed arguments. (I) At each pixel of the diffuse strain band, slip is predominantly ascribed to only one of the possible 14 slip systems, as shown in subplots (b\textsubscript{1,4}) and (c\textsubscript{1,8,10}) (note that the slip activity in subplots (c\textsubscript{5,6}) is outside of the strain band), in line with our general knowledge on plasticity, even though the \SI\ method is free to identify slip activity on many slip systems simultaneously. (II) The slip systems identified by \SI\ in each variant describe the displacement gradients from experiments in each pixel accurately, i.e. the residual between $\mathbf{H}^{exp}$ and $\mathbf{H}^{thresh}$ is below or at the noise level. (III) In all four variants through which the strain band passes, each with a very different crystal orientation, \SI\ has automatically identified slip activity on a slip system whose slip trace aligns with the diffuse strain band direction, providing an independent validation. (IV) The similarity between the in-plane slip traces and directions for the active slip systems from different variants suggests slip transfer, as is expected for a diffuse strain band.

A limitation of the \SI\ method is that $\mathbf{H}^{exp}$ is a 2D tensor, since no 3D deformation data is measured. Therefore, slip systems whose 2D slip trace and direction are similar to each other can be difficult to distinguish for some cases. Furthermore, it must be noted that \SI\ identification can be performed for any of the deformation increments for which noticeable displacements are recorded ($\varepsilon_{vm} > 0.02$). However, \SI\ identification is only performed for the first increment of noticeable plasticity, since later increments get increasingly more affected by rotation, warping and damage in the microstructure, thereby reducing the reliability of the \SI\ method. It is thereby assumed that activated slip systems in the first noticeable strain increment stay active during further deformation.

\subsection{Simulations}
\label{sec:app_sim}

The \cps\ model follows the large deformation theory in which the deformation gradient tensor $\mathbf{F}$ is decomposed into elastic ($\mathbf{F}_{\mathrm{e}}$) and plastic ($\mathbf{F}_{\mathrm{p}}$) contributions:

\begin{equation}
\mathbf{F} = \mathbf{F}_{\mathrm{e}}\mathbf{F}_{\mathrm{p}}
\ . \
\end{equation}
The rate of plastic deformation is described by a plastic velocity gradient tensor $\mathbf{L}_{\mathrm{p}}$:
\begin{equation}
\mathbf{L}_{\mathrm{p}}=\sum_{\alpha=1}^{n} \dot{\gamma}^{\alpha} \Vec{s}^{\:\alpha} \otimes \Vec{n}^{\alpha},
\end{equation}
where $\dot{\gamma}^{\alpha}$ indicates the shear rate on slip system $\alpha$.
A standard phenomenological approach is employed in which the shear rate on each slip system is determined by a resolved shear stress $\tau^{\alpha}$ and a flow resistance $s^{\alpha}$ \cite{Roters2019}. 
The shear rate for slip system $\alpha$ is calculated as:
\begin{equation}
\dot{\gamma}^{\alpha}=\dot{\gamma}_{0}\left|\frac{\tau^{\alpha}}{s^{\alpha}}\right|^{\frac{1}{m}} \operatorname{sgn}\left(\tau^{\alpha}\right) ,
\end{equation}
where $\dot{\gamma}_{0}$ is a reference shear rate and $m$ is a rate sensitivity parameter. Hardening is implemented as an evolution of the flow resistance \cite{Bronkhorst1992PolycrystallineMetals}:
\begin{equation}
\dot{s}^{\alpha}=\sum_{\beta=1}^{n} h^{\alpha \beta}\left|\dot{\gamma}^{\beta}\right|
\ , \
h^{\alpha \beta}=h_{0}\left(1-\frac{s^{\alpha}}{s_{\infty}}\right)^{a}\left(q+(1-q) \delta^{\alpha \beta}\right) ,
\end{equation}
wherein $h_0$ is the initial hardening, $a$ the hardening shape factor, $\delta^{\alpha \beta}$ the Kronecker delta, and $s^{\alpha}$ the slip resistance, which evolves from an initial value $s_{0}$ to a saturation value $s_{\infty}$. $h^{\alpha \beta}$ is referred to as the hardening modulus which evolves due to self-hardening of slip system $\alpha$ and latent hardening induced by other slip systems $\beta$. The latent hardening parameter $q$ describes the ratio between self-hardening and latent hardening. 

The \cpm\ model differs from the \cps\ in the contributions to the plastic velocity gradient tensor $\mathbf{L}_{\mathrm{P}}$ for the martensite phase, with the \cps\ model employing
\begin{equation} \label{eq:velgrad1}
\begin{aligned}
\mathbf{L}_{\mathrm{P}} = \mathbf{L}_{\mathrm{P}}^{\alpha^{\prime}},
\end{aligned} 
\end{equation}
while for the \cpm\ model, we employ the rule of mixture,
\begin{equation} \label{eq:velgrad2}
\begin{aligned}
 \mathbf{L}_{\mathrm{P}} = \varphi \mathbf{L}_{\mathrm{P}}^{\gamma} + (1-\varphi)\mathbf{L}_{\mathrm{P}}^{\alpha^{\prime}} , 
\end{aligned}
\end{equation}
in which $\alpha'$ and $\gamma$ refer to the martensite and austenite phase and $\varphi$ refers to the volume fraction of austenite, which is taken at 5\% for this work. {This value is based on measurements on a variety of interlath-retained austenite containing steel grades, wherein the volume fraction ranged from 1\% to 8\% \cite{thomas1977retained, SAMUEL1985, sherman2007, YUAN2012}. A parameter study of this value in our simulations gave no significant differences in results.} While $\mathbf{L}_{\mathrm{P}}^{\gamma}$ includes contributions from all FCC austenite slip systems, the model is implemented by only including the three $\{111\}\langle110\rangle_{\gamma}$ FCC austenite slip systems, placed along the $(011)_{\alpha'}$ habit plane of each variant. Due to lath morphology and the existing OR between $\gamma$ and $\alpha'$, the slip activity of OHP slip systems in austenite films are constrained by the ones from lath martensite, and hence, will not contribute to the softer habit plane sliding mechanism. For the detailed implementation of the \cpm\ model, the reader is referred to the work of Rezazadeh \textit{et al.} \cite{VahidP4}. In both \cpm\ and \cps\ models, 12 $\{110\}\langle111\rangle$ slip systems are used for BCC martensite. For ferrite, the \cps\ model with 12 BCC slip systems is used in all cases. 

\subsubsection{Microstructure model}
\label{sec:app_microstructure}

Experimental data is taken as input into the simulation framework. The grid structure used in DAMASK to describe the geometry of a Representative Volume Element (RVE) allows for EBSD data (crystal orientations and phases) to be directly translated into numerical simulations, which means no meshing is required. Variant numbers from the PAG reconstruction are used in the \cpm\ model to assign local habit plane orientations employing the KS OR \cite{VahidP4}. For the simulations to be computationally feasible, a grid spacing of 40 nm is used, for which the data has been coarsened from the 20 nm pixel size on which the data is aligned. Other microstructural information, such as variant orientations and phase data, have also directly been inserted from experiments.

\begin{figure}[H]
    \centering
    \includegraphics[width=\linewidth]{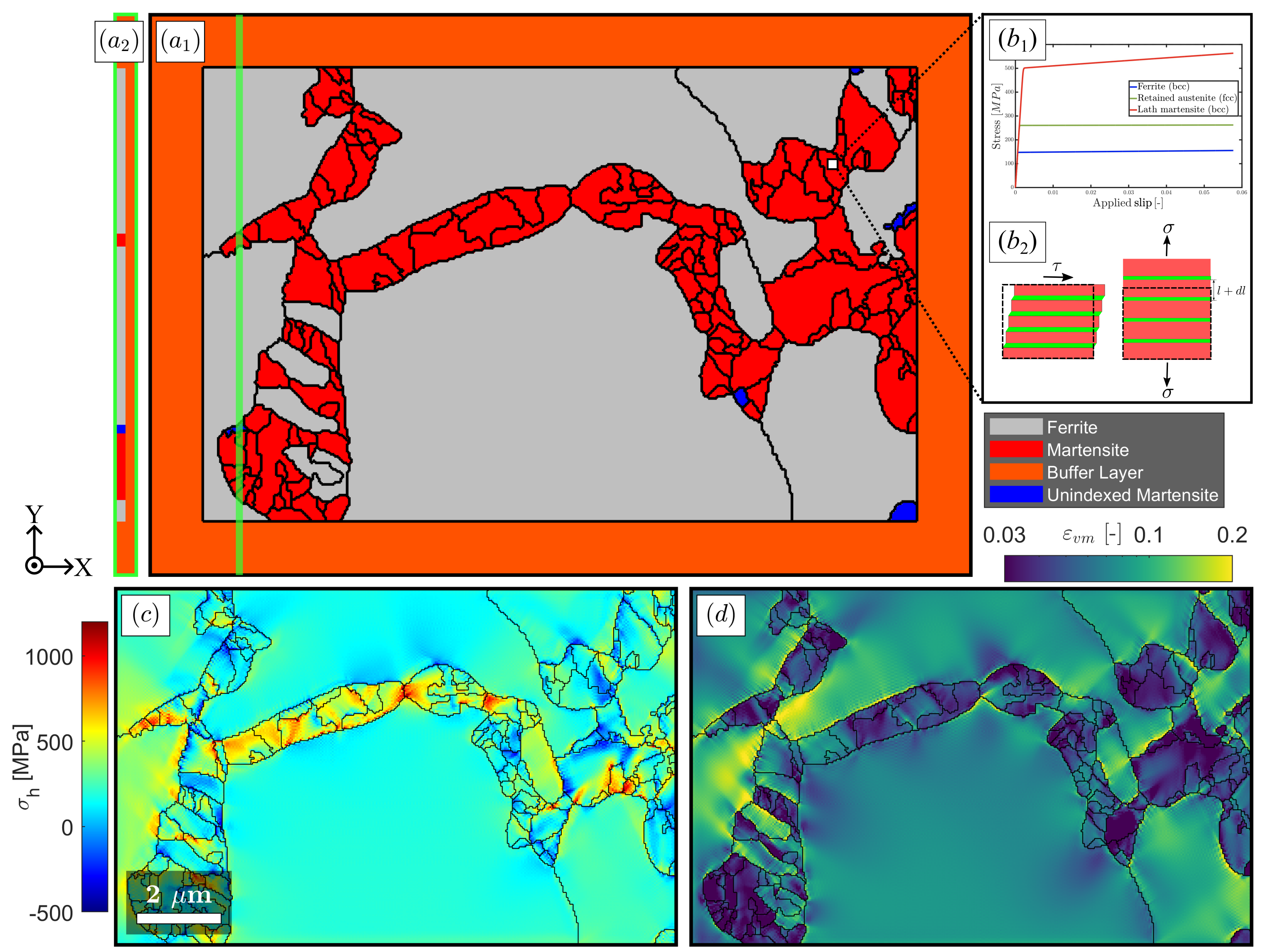}
    \caption{Overview of the simulation microstructure and results when using the \cpm\ model. (a\textsubscript{1}) The simulated region, with martensite and ferrite in red and gray, and the surrounding and underlying buffer layers marked in orange. The vertical green line indicates the location of the cross-section displayed in (a\textsubscript{2}). (b) Schematic view and properties of the "matrix-film" (martensite-austenite) model which is represented in each martensite voxel. (b\textsubscript{1}) Stress-strain behaviour of a single slip system in ferrite, martensite and austenite. (b\textsubscript{2}) The deformation for an applied shear stress $\tau$ and a normal stress $\sigma$ drawn schematically for the laminate model \cite{VahidP4}. (c-f) Simulation results: (c) hydrostatic stress and (d) 2D equivalent true strain.}
    \label{fig:methodSim}
\end{figure}

Where possible, simulations are performed using a larger microstructure than the region tracked by SEM-DIC, to minimize artefacts inside the relevant region, caused by the unrealistic periodic boundary conditions employed in the FFT-based solver. Since periodicity artefacts generally only influence the edges of the RVE, and the direct surrounding microstructure of an ROI has been found to be most decisive for stress/strain partitioning, periodic repetition of the microstructure is a valid assumption \cite{TASAN2014386,Tasan2014}. Because some experimental regions are fairly close to the boundary of the ROI, the ROI size is extended 20\% in all in-plane directions with an isotropic buffer layer, in order to minimize periodicity artefacts in the microstructure. Based on the global ferrite-martensite volume fraction, volume-averaged yield and hardening properties of martensite (20\%) and ferrite (80\%) are used for the isotropic properties. In Z-direction, out of the sample plane, the microstructure is repeated over 5 element layers, underneath which 5 isotropic buffer element layers are placed. This underlying buffer layer ensures that slip systems whose slip directions are (strongly) tilted with respect to the sample plane are not fully constrained by the out-of-plane periodic boundary conditions, but instead are able to localize relatively freely, similar to reality \cite{VahidP4}. An example of the simulation, based on the microstructure from \autoref{fig:fig1method}, is shown in \autoref{fig:methodSim}(a\textsubscript{1}), which includes a cross-section showing the buffer layer (in orange) in the Z direction (a\textsubscript{2}). The martensite-austenite laminate model, which is implemented in each martensite voxel, is vizualized in \autoref{fig:methodSim}(b), showing the properties of a single slip system in each phase plotted in (b\us{1}) and a schematic view of the deformations in (b\us{2}) \cite{VahidP4}. 

\subsubsection{Material Properties}
\label{sec:app_matProp}
The material parameters employed to model the response of the phases Martensite, Austenite and Ferrite are taken from Maresca \textit{et al.} \cite{maresca2016reduced} and shown in \autoref{tab:Table1}. In particular, the phase contrast between the soft FCC mechanism embedded in the habit plane and the lath martensite BCC crystal is $\sim 2$ \cite{maresca2016reduced,VahidP4}. A parameter study was performed and showed that the specific value for the phase contrast does not significantly influence the qualitative results of the simulations. 

\begin{table}[H]
\caption{Material parameters of phases in \cpm\ and \cps\ models. \cite{maresca2016reduced,VahidP4}}
\centering
\begin{tabular}{llllll}
Parameter                   & Symbol    & M Lath        & A Film        & F      & Buffer  \\ 
\addlinespace
\hline
\addlinespace
Elasticity Component 11 [GPa]     & $C_{11}$       & 283       & 283    & 283      & 283\\
Elasticity Component 12 [GPa]     & $C_{12}$       & 121  & 121    & 121       & 121  \\
Elasticity Component 44 [GPa]     & $C_{44}$       & 81      & 81      & 81       & 81   \\
Reference Slip Rate [$s^{-1}$]  & $\dot{\gamma}_0$ & 0.001   & 0.001  & 0.001  & 0.001  \\
Initial Slip Resistance [GPa]     & $s_0$        & 0.51  & 0.265     & 0.15     & 0.22 \\
Saturation Slip Resistance [GPa]  & $s_\infty$      & 2 & 0.34  & 0.25 & 0.6  \\
Reference Hardening Modulus [GPa] & $h_0$        & 1.5  & 0.25    & 0.5  & 0.7 \\
Strain Rate Sensitivity [-]     & $m$         & 0.05  & 0.05  & 0.05& 0.05 \\
Hardening Exponent [-]         & $a$         & 1.5   & 1.5 & 1.5  & 1.5\\
Latent/Self Hardening Ratio [-] & $q$         & 1.4    & 1.4      & 1.4    & 1.4  \\
Taylor Factor [-] & $M$ & - & - & - & 2.4 \\
Slip Family         & $\{\Vec{n}\}$   & $\{110\}_{\alpha'}$ & $\{111\}_{\gamma}$ & $\{110\}$ & -  \\
       & $\langle\Vec{s}\rangle$   & $\langle111\rangle_{\alpha'}$ & $\langle110\rangle_{\gamma}$ & $\langle111\rangle$ & -  \\
Number of Slip Systems        &  $n$  & 12 & 3 & 12 & -  \\
\addlinespace
\hline
\end{tabular}
\label{tab:Table1}
\end{table}

\subsubsection{Boundary Conditions}
\label{sec:app_BC}

The loading conditions are the same for all simulation results, and are defined in terms of a volume-averaged deformation rate tensor $\dot{\overline{\mathbf{F}}}$ and a first Piola-Kirchoff stress tensor $\mathbf{P}$ \cite{Roters2019}. The applied boundary conditions are:

\begin{equation}
\dot{\overline{\mathbf{F}}}=\left[\begin{array}{ccc}
\dot{\lambda} & * & * \\
0 & * & * \\
0 & 0 & *
\end{array}\right], \overline{\mathbf{P}}=\left[\begin{array}{lll}
* & 0 & 0 \\
* & 0 & 0 \\
* & * & 0
\end{array}\right],
\end{equation}

where $\overline{\square}$ indicates a volume-averaged property, $*$ indicates entries that are free to evolve and $\dot{\lambda}$ is the stretch rate per increment, which is always defined so that the component $\dot{\overline{\mathbf{F}}}_{11}$ is 1.1 at the end of the simulation, corresponding to $\overline{\varepsilon}_{xx}\sim= 9.5 \%$, after which the increment is chosen which corresponds best to the experimental mean strain value.

\subsubsection{Example of Simulation results}

Examples of the most relevant results, namely, 2D equivalent true (logarithmic) strain $\varepsilon_{vm}$ and hydrostatic stress $\sigma_{h}$ are shown (for the \cpm), in \autoref{fig:methodSim}(c-f), including grain boundaries. The 2D equivalent true strain is employed to allow for a better comparison to experimental strain maps.

\centering
\noindent\rule{8cm}{0.4pt}

\clearpage



\end{document}